%% file: paper.tex
\documentclass[acmsmall]{acmart}

\setcopyright{none}

\usepackage{amssymb}
\usepackage{array}
\usepackage{hyperref}
\usepackage{subcaption}
\usepackage[labelfont=bf]{caption}
\usepackage[noline,ruled,commentsnumbered,linesnumbered,titlenumbered]{algorithm2e}
\usepackage{etoolbox} % automatic table row numbering
\usepackage{tablefootnote}
\usepackage{multirow}
\usepackage{colortbl}
\usepackage{xcolor}
\usepackage{footnote}
\makesavenoteenv{tabular}

\newcommand{\cal}{\mathcal} % if \cal is unrecognized
\newcolumntype{x}[1]{>{\centering\arraybackslash\hspace{0pt}}m{#1}} % more compact table columns
\newcounter{magicrownumbers} % automatic table row numbering

\begin{document}

\title{Efficient Online Classification and Tracking on Resource-constrained IoT Devices}
\thanks{$^\ast$This is a substantially enhanced version of prior papers \cite{aftabAPsys, osplug}. This paper is to be published in ACM Transactions on Internet of Things (TIOT)
}

\author{Muhammad Aftab}
\affiliation{%
  \institution{Aalborg University, Denmark}
}
\email{muhafab@cs.aau.dk}

\author{Sid Chi-Kin Chau}
\affiliation{%
  \institution{Australian National University, Australia}
}
\email{sid.chau@anu.edu.au}

\author{Prashant Shenoy}
\affiliation{%
  \institution{University of Massachusetts, Amherst, USA}
}
\email{shenoy@cs.umass.edu}

\begin{abstract}
\input{abstract}

\end{abstract}

%\begin{CCSXML}
%  <ccs2012>
%    <concept>
%    <concept_id>10010520.10010553.10010562.10010563</concept_id>
%    <concept_desc>Computer systems organization~Embedded hardware</concept_desc>
%    <concept_significance>500</concept_significance>
%    </concept>
%    <concept>
%    <concept_id>10010520.10010553.10010562.10010564</concept_id>
%    <concept_desc>Computer systems organization~Embedded software</concept_desc>
%    <concept_significance>500</concept_significance>
%    </concept>
%    <concept>
%    <concept_id>10010583.10010588.10010596</concept_id>
%    <concept_desc>Hardware~Sensor devices and platforms</concept_desc>
%    <concept_significance>300</concept_significance>
%    </concept>
%  </ccs2012>
%\end{CCSXML}
\ccsdesc[500]{Computer systems organization~Embedded hardware}
\ccsdesc[500]{Computer systems organization~Embedded software}
\ccsdesc[300]{Hardware~Sensor devices and platforms}

\setcopyright{acmlicensed}
\acmJournal{TIOT}
\acmYear{2020} \acmVolume{1} \acmNumber{1} \acmArticle{1} \acmMonth{1} \acmPrice{15.00}\acmDOI{10.1145/3392051}

\keywords{Smart power plugs, Internet-of-things, online information processing, resource-constrained systems}

\maketitle

\input{intro}
\input{background}

\input{model}

\input{event}
\input{track}

\input{testbed}

\input{eval}
\input{related}
\input{concl}
\appendix

\bibliographystyle{ACM-Reference-Format}
%\bibliographystyle{plain}
%\bibliographystyle{abbrv}
%\scriptsize
\bibliography{refs}

\section*{Acknowledgements} 
We thank the anonymous reviewers and associate editor Valerie Issarny for their helpful comments. Sid Chau was supported in part by ARC Discovery Project GA69027.  Prashant Shenoy was supported in part by NSF grants 1505422 and 1645952.

\newpage
\textbf{APPENDIX}
\input{append}

\end{document}

%% file: abstract.tex
Timely processing has been increasingly required on smart IoT devices, which leads to directly implementing information processing tasks on an IoT device for bandwidth savings and privacy assurance. Particularly, monitoring and tracking the observed signals in continuous form are common tasks for a variety of near real-time processing IoT devices, such as in smart homes, body-area and environmental sensing applications. However, these systems are likely low-cost resource-constrained embedded systems, equipped with compact memory space, whereby the ability to store the full information state of continuous signals is limited. Hence, in this paper$^\ast$ we develop solutions of efficient timely processing embedded systems for online classification and tracking of continuous signals with compact memory space. Particularly, we focus on the application of smart plugs that are capable of timely classification of appliance types and tracking of appliance behavior in a standalone manner. We implemented a smart plug prototype using low-cost Arduino platform with small amount of memory space to demonstrate the following timely processing operations: (1) learning and classifying the patterns associated with the continuous power consumption signals, and (2) tracking the occurrences of signal patterns using small local memory space. Furthermore, our system designs are also sufficiently generic for timely monitoring and tracking applications in other resource-constrained IoT devices.

%% file: intro.tex
\section{Introduction} \label{sec:smartplug:intro}

The rise of IoT systems enables diverse monitoring and tracking applications, such as smart sensors and devices for smart homes, as well as body-area and environmental sensing.  In these applications, special system designs are required to address a number of common challenges. First, IoT systems for monitoring and tracking applications are usually implemented in low-cost resource-constrained embedded systems, which only allow compact memory space, whereby the ability to store the full information state is limited. Second, timely processing has been increasingly required on smart IoT devices, which leads to implementing near real-time information processing tasks as close to the end users as possible, for instance, directly implementing on an IoT  device for bandwidth savings and privacy assurance. Hence, it is increasingly important to put basic timely computation as close as possible to the physical system, making the IoT devices (e.g., sensors, tags) as ``smart'' as possible. However, it is challenging to implement timely processing tasks in resource-constrained embedded systems, because of the limited processing power and memory space. 

To address these challenges, a useful paradigm is {\em streaming data} (or data streams) processing systems \cite{datastream_survey}, which are systems considering a sequential stream of input data using a small amount of  local memory space in a standalone manner. These systems are suitable for timely processing IoT systems with constrained local memory space and limited external communications. However, traditional settings of streaming data inputs often consider discrete digital data, such as data objects carrying certain unique digital identifiers. On the other hand, the paradigm of timely processing IoT, which aims to integrate with physical environments \cite{insitusensnet}, has been increasingly applied to diverse applications of near real-time monitoring and tracking on the observed signals in continuous form, such as analogue sensors for physical, biological, or chemical aspects. 

For example, one application is the smart plugs, which are computing devices augmented to power plugs to perform monitoring and tracking tasks on continuous power consumption signals, as well as inference and diagnosis tasks for the connected appliances. Smart plugs are usually embedded systems with constrained local memory space and limited external communications. Another similar application is body-area or biomedical sensors that track and infer continuous biological signals. Note that this can be extended to any processing systems for performing timely sensing, tracking and inference tasks with continuous signals.

\begin{figure*}[htb!]
    \centering 
    \includegraphics[width=\textwidth]{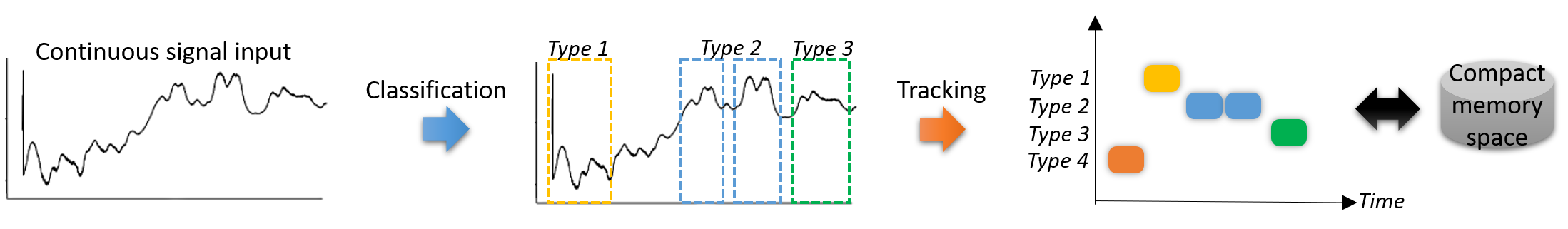}
    \caption{Basic framework of a system for timely classification and tracking of continuous signals, using compact local memory space.}
    \Description{Basic framework of a system for timely classification and tracking of continuous signals, using compact local memory space.}
    \label{fig:framework}
\end{figure*}

In this paper, we consider timely processing IoT systems that are able to classify and record the occurrences of signal patterns over time. Also, the data of signal patterns will be useful to identify temporal correlations and the context of events. For example, the activities of occupants can be identified from the signal patterns in smart home applications. This paper studies the problems of efficient tracking of occurrences using small local memory space. We aim to extend the typical streaming data processing systems to consider continuous signals. The basic framework of such a tracking and monitoring system is illustrated in Fig.~\ref{fig:framework}.

In summary, we study tracking and monitoring systems to support that following functions:
{\color{black} 
\begin{enumerate}
  \item Timely learning and classifying patterns of continuous signals from known classes of signal patterns.
  \item Timely learning and classifying  unknown patterns of continuous signals.
  \item Timely tracking occurrences of signal patterns of interests using small local memory space.
\end{enumerate}
}

In particular, we focus on the application of smart plugs, which can provide a practical testbed for evaluating the tracking and monitoring system solutions. We developed standalone smart plugs that are capable of timely classification of appliance types and tracking of appliance behavior in a standalone manner. We built and implemented a smart plug  prototype using low-cost Arduino platform with a small amount of memory space. Nonetheless, our system designs are also sufficiently generic for other timely monitoring and tracking applications of continuous signals.

The rest of the paper is organized as follows. Section~\ref{sec:smartplug:background} provides a review of the relevant background. We formulate the problems of timely pattern classification and occurrence tracking that are implemented in the smart plug system in Section~\ref{sec:smartplug:problem}. The techniques and algorithms of timely pattern classification and occurrence tracking are provided in Sections~\ref{sec:smartplug:pattern_classif} and \ref{sec:smartplug:pattern_track}, respectively. We present a prototype implementation of a smart plug system in Section~\ref{sec:smartplug:testbed}, and its detailed experimental evaluation study in Section~\ref{sec:smartplug:eval}. A survey of related work is provided in Section~\ref{sec:smartplug:related}. Finally, we conclude the paper and discuss future extensions in Section~\ref{sec:smartplug:concl}. {\color{black} A table of key notations and their definitions is provided in Appendix.}

%% file: background.tex
\section{Background} \label{sec:smartplug:background}

Efficient tracking of continuous signals over time is critical to several cyber-physical systems. In this work, we adopt streaming data algorithms for tracking the occurrences of signal patterns in embedded systems. There have been extensive theoretical studies of streaming data algorithms for counting  problems (e.g., heavy hitters \cite{Cormode_heavyhitter}, frequency moments \cite{superspreader,flowentropy,bitmap}).  In the past, the applications of these studies are related to database systems and network traffic measurement.  In this paper, we focus on the specific application to cyber-physical systems for tracking continuous signals. A novel contribution of this work is to implement streaming data algorithms for tracking the occurrences in embedded systems. The implementation of streaming data algorithms in embedded systems faces several new challenges, as these systems can only handle simple data processing, without sophisticated mathematical processing capability. 

In this work, a smart plug prototype is developed for the classification of appliance types and behavior. Traditionally, the classification tasks for appliances can be achieved using either non-intrusive methods \cite{nilm_survey} or intrusive methods. Non-intrusive methods involve passive measurements of power consumption of the entire apartment and disaggregating the data to identify individual appliances, whereas intrusive methods require active measurements of power consumption of individual appliances using  special power meters or smart plugs. Our smart plug prototype employs intrusive methods. Note that most of the prior studies of appliance classification were carried out by simulations using offline data collection. Few recent studies have implemented in embedded systems.  {\color{black} For example, a system prototype is presented in \cite{auto_plug}, called AutoPlug, which can perform appliance classification and identification on a wireless gateway connecting to multiple power plugs. Unlike our smart plug prototype using resource-constrained Arduino platform, AutoPlug is based on more powerful Raspberry Pi 2 platform.} 

Overall, there are several differences between our study and the prior studies as follows.
\begin{enumerate}

\item Most of the proposed non-intrusive, as well as intrusive methods, do not perform near real-time appliance identification and instead deal with pre-recorded data in an offline manner. Our smart plug, on the other hand, is able to perform timely monitoring and classification of appliances using only the data seen thus far, without any knowledge of the future data.

\item Some recent efforts have also explored near real-time appliance identification. However, there are several fundamental differences with this work. For example, we focus on resource-constrained embedded system platform and online algorithms. See Section~\ref{sec:smartplug:related} for a detailed comparison.

\item Our smart plug is developed to support efficient streaming data algorithms, unlike other simple smart plug projects. It can perform advanced processing tasks locally in an online manner using small memory space, which requires very efficient implementation of the processing algorithms. 

\item  The existing classification and tracking algorithms are not designed for embedded systems in mind. Thus, we make non-trivial modifications to the existing algorithms so that it can be implemented in smart plug using limited processing power and memory size.
      
\end{enumerate}

Furthermore, the contributions of this paper are summarized as follows.

\begin{enumerate}

\item We devised techniques and algorithms to perform timely classification of continuous
signals (as opposed to the classification of discrete data in the extant literature). Our techniques are efficient to run on low-cost embedded systems.

\item We designed algorithms based on streaming data to perform timely occurrence tracking
using compact memory space in low-cost embedded systems, which is novel in traditional streaming data literature.

\item We implemented a hardware prototype of a smart plug and implemented all of the proposed
techniques on the smart plug to process power consumption signals in a timely manner. Our hardware prototype demonstrated practical effectiveness of the methods in real-world applications of smart plugs, which has been demonstrated in limited extent in the state of the art.

\item We conducted detailed evaluations of our techniques and prototype. Our evaluations show
that our system can classify and track with good accuracy in compact memory space in real-world applications of smart plugs.

\end{enumerate}

%% file: model.tex
\section{Problem Formulation} \label{sec:smartplug:problem}

%In this section, we formulate the problems of pattern classification, occurrence tracking, {\color{red} and correlation tracking} that are implemented in the smart plug prototype. 
In this section, we formulate the problems of pattern classification and occurrence tracking that are implemented in the smart plug prototype. {\color{black} Note that a table of key notations and their definitions is provided in Appendix.}

For generality, we first describe a generic setting of sensors that monitor and track certain continuous signals. We then apply to the specific setting of smart plugs. %In the following, we provide the basic ideas of signal classification and occurrence tracking.

\subsection{Signal Classification} \label{subsec:prob:classif}

We denote by $x$ a stream of continuous signals observed by a sensor. For convenience, we assume slotted time, such as $(x_1, x_2, x_3, ... )$. The signals may be triggered by various events at different times that are not revealed to the sensor. Given the continuous nature of signals, the sensor needs to interpret and identify the states and transitions embodied by the signals. 
%We note that there may be also the presence of noise or measurement errors that affect the measurements.

To classify the signal patterns, we first detect a proper segmentation of continuous signals, which captures the passages and terminations of specific patterns. Note that patterns of signals do not necessarily occur at a fixed time interval. In this work, segmentation is based on the detection of state changes, where signals can exhibit different characteristics at different times. For example, the consumption of appliances may exhibit various state transitions, such as different operation cycles of a washing machine. Next, segments of signals will be classified according to the following two approaches.

\subsubsection{Known Classes of Signal Patterns}

The signals may be triggered by a finite number of possible classes of signal patterns. The list of possible classes of signal patterns may be known by the sensor in a-prior. For example, signals exhibiting certain common stochastic properties can be triggered by a common class of signal patterns.  Regarding smart plugs, there are common stochastic models suitable for describing the power consumption patterns of appliances, which can be used as the prior known knowledge for classifying power consumption signals.

\subsubsection{Unknown Signal Patterns}

Alternatively, one can employ a clustering algorithm to classify the signal patterns in a way to minimize the discrepancy in each cluster. Let $x[t_1:t_2]$ be the segment of signal from time $t_1$ to $t_2$, and ${\cal S}$ be a non-overlapping segmentation of time, such that if $(t_1, t_2), (t_3, t_4) \in {\cal S}$, then $x[t_1: t_2]$ and $x[t_3: t_4]$ are non-overlapping segments of $x$. We denote the set of clusters of patterns by ${\cal C}$. For each $i \in {\cal C}$, denote $c_i$ be a canonical signal that is selected to represent a cluster of patterns. Let ${\sf d}\big(x[t_1: t_2], c_i\big)$ be a distance metric between $x[t_1: t_2]$ and $c_i$. One example of distance metric is $\ell_2$ norm that measures the total absolute discrepancy over the designated interval. 

We aim to find suitable ${\cal S}$ and ${\cal C}$, as to minimize  an objective of the following general form:
\begin{equation}\label{eq:unkown_patterns}
\min_{{\cal S}, {\cal C}} \ \rho |{\cal C}| + \sum_{i \in {\cal C}} \sum_{(t_1, t_2) \in {\cal S}_i} {\sf d}\big(x[t_1, t_2], c_i\big)
\end{equation}
where $\rho$ is a weight parameter, and ${\cal S}_i$ is the set of segments belonging to the $i$-th cluster, defined as follows:
\begin{equation}
\begin{array}{@{}l@{}}
{\cal S}_i \triangleq \Big\{(t_1, t_2) \in {\cal S} \mid {\sf d}\big(x(t_1, t_2), c_i\big) \le {\sf d}\big(x(t_1, t_2), c_j\big), \forall j \in {\cal C}\backslash \{ i \} \Big\},
\end{array}
\end{equation}
In particular, we consider $k$-mean clustering, such that  the number of clusters is at most $k$, where $|{\cal C}| \le k$, namely, minimizing the following objective:
\begin{equation}\label{eq:kmin_patterns}
\min_{{\cal S}, {\cal C}: |{\cal C}|  \le k} \ \sum_{i \in {\cal C}} \sum_{(t_1, t_2) \in {\cal S}_i} {\sf d}\big(x(t_1, t_2), c_i\big).
\end{equation}
An optimal solution to the above objective function will yield a segmentation of the input signal into a set of clusters where similar signal patterns are classified into the same cluster.

\subsection{Occurrence Tracking} \label{subsec:prob:track}

After classifying the signals, we can track the occurrences of each pattern over time. A detailed tracking history may require large memory space, in particular when the time epoch \footnote{\color{black} An epoch is a particular period of time marked by distinctive features or events.} is long, or the number of possible patterns is large. Hence, we focus on the notion of {\em approximate tracking}.

In general, there is a stream of items with multiple occurrences. We want to record the items with the most prominent occurrences when observing the stream continuously. Note that we do not know the number of distinct items in advance, and we want to use memory space much less than the number of distinct items in the stream.

Suppose that the set of observed patterns are ${\cal C}$.  Note that ${\cal C}$ may be a large set. The occurrences of each pattern will be recorded over {\color{black}  a limited time horizon ${\cal T}$.} Let $N_i^{\cal T}$ be the true total number of occurrences of pattern $i \in {\cal C}$ {\color{black}  in the time horizon ${\cal T}$.} Our objective is that given a memory size ${\sf c}$, one can identify the occurrences of the common patterns with high accuracy relative to the length of epoch $| {\cal T}|$.  Note that the difficulty is the memory size ${\sf c}$ is a fixed constant that cannot grow at the same rate as $|{\cal C}|$. 

In approximate tracking, we aim to obtain an estimated total number of occurrences $\widehat{N}_i^{\cal T}$ at the end of ${\cal T}$. We are interested in an assured bound of error probability for accuracy guarantee from true total number of occurrences, called ($\epsilon,\delta$)-accurate estimation, which satisfies:
\begin{equation}
\displaystyle {\mathbb P}\Big(|{\widehat{N}}_i^{\cal T} - N_i^{\cal T}| \ge \epsilon \cdot |{\cal T}|\Big) < \delta,
\end{equation}
where $\epsilon$ and $\delta$ are the parameters for controlling the trade-off between accuracy and memory size.

For smart plugs, we will record the occurrences of power consumption patterns in multiple intervals of time, for example, the last hour, the last day, the last 2 days, the last week, etc. In practice, the occurrences will be recorded in a rolling window fashion. Due to the limited memory space, some occurrences in different intervals will be aggregated over time. For example, we will merge the data of the last seven days into one single aggregate record for the last week. Hence, approximate tracking is required to support the merging of data.

%% file: event.tex
\section{Classifying Continuous Signals}\label{sec:smartplug:pattern_classif}

This section presents the basic ideas, techniques, and algorithms for learning and classifying the patterns associated with continuous signals in an online manner. Although the ideas can be applied to generic applications of continuous signal monitoring and tracking, we particularly consider the setting of continuous power consumption signals of electrical appliances.

\subsection{Pattern Detection} \label{sec:smartplug:pattern_dect}

The first step is the detection of the patterns embodied by the continuous signals. For power consumption signals, an appliance may transit through several operating states, in which the power consumption patterns usually vary from one state to the next. A state transition may be characterized by multiple components\footnote{For example, a dishwasher contains both a motor and a heating element. The motor powers the pump to propel and sprays hot water on the dishes. The heating element is responsible for heating the water for washing or heating the air for drying. Similarly, an air-conditioner contains both a compressor and a fan, among other basic functions.}. The same state transition may also occur at several discrete levels of the input signals\footnote{For instance, an electric iron is equipped with a temperature control dial, which allows a user to select the iron's operating temperature. Adjusting the control dial will cause the iron to change its power consumption level.}.
We aim to identify such state transition patterns in the continuous signals. More specifically, we propose an online algorithm to detect state transition patterns by analyzing the continuous data stream. 
%Our algorithm is an online variant of Algorithm~\ref{alg:scd_offline}, which is an offline algorithm originally proposed in \cite{nilm}.

\subsubsection{Offline Detection} \label{sec:smartplug:offline_dect}

We first describe how patterns can be detected from the recorded data in an offline manner. This algorithm is proposed in \cite{nilm}. It is based on the concept of \emph{Approximate Entropy} ({\tt ApEn}) and {\em Canny Edge Detection}.  {\tt ApEn} is a metric for estimating the repeatability or predictability of a time series \cite{ApEn}. It was originally developed for heart rate analysis and later is applied to a wide range of applications such as psychology and financial analysis. {\tt ApEn} of a time series $(x_i)_{i=1}^N$ is computed in the following steps:

\begin{enumerate}
  \item Fix two parameters: a positive integer $m$ and a positive real number $\theta$, where $m$ represents the length of a sub-sequence, and $\theta$ is the similarity threshold between a pair of sub-sequences.

  \item Consider the set of sub-sequences of length $m$ in $x$, defined by $s^m = \big\{x[i:i+m-1] \big\}_{i= 1}^{N-m+1}$.

  \item  For each $i$, define $C_i^m(\phi)$ as the fraction of sub-sequences in $s^m$ that are similar to $x[i:i+m-1]$. Two sub-sequences, $x[i:i+m-1]$ and $x[j:j+m-1]$, are similar if the difference between any pair of the corresponding values in the sub-sequences is less than or equal to $\theta$, namely,
        \begin{equation}
          |x_{i+k} - x_{j+k}| \le \theta ,\quad  \mbox{for all\ } k\in \{1,...,m\}.
        \end{equation}

  \item Finally, ${\tt ApEn}$ is computed as follows:
        \begin{equation}
          {\tt ApEn}((x_i)_{i=1}^N) = \frac{1}{N-m+1} \sum_{i=1}^{N-m+1} \log C_i^m - \frac{1}{N-m} \sum_{i=1}^{N-m}  \log C_i^{m+1}.
        \end{equation}

\end{enumerate}

${\tt ApEn}$ can be used to measure the regularity in power consumption signals. Specifically, if similar power consumption patterns are similar, then {\tt ApEn} will be small. Conversely, if there are irregular patterns, then {\tt ApEn} will be large.

\begin{minipage}[t]{0.470\linewidth}
  \begin{algorithm}[H]
    \caption{{\tt OFLStateTrans}}
    \label{alg:scd_offline}
    \SetAlgoLined
    \DontPrintSemicolon
    \SetKwInput{Input}{Input}
    \SetKwInOut{Output}{Output}
    \SetKwInOut{Initialization}{Initialization}
    \SetKwComment{Comment}{$\triangleright$\hspace{-4pt}\ }{}\SetInd{0.2em}{0.35em}
    \Input{$x,\ \phi,\ \omega$}
    \ForAll{$i  \in \{ 1, ..., |x|-\phi\}$}{
      $H_i \leftarrow  {\tt ApEn}(x[i:i+\phi])$
    }
    $E' \leftarrow {\tt CannyEdge1D}\big((H_i)_{i=1}^{|x|-\phi}\big) $ \\
    $E \leftarrow {\tt RemoveCloseEdges}(E', \omega) $ \\
    {\bf return} $E$
  \end{algorithm}
\end{minipage}
\quad
\begin{minipage}[t]{0.470\linewidth}
  \begin{algorithm}[H]
    \caption{{\tt ONLStateTrans}}
    \label{alg:scd_online}
    \SetAlgoLined
    \DontPrintSemicolon
    \SetKwInput{Input}{Input}
    \SetKwInOut{Output}{Output}
    \SetKwInOut{Initialization}{Initialization}
    \SetKwComment{Comment}{$\triangleright$\hspace{-4pt}\ }{}\SetInd{0.2em}{0.35em}
    \Initialization{$E \leftarrow \varnothing$}
    \Input{$x,\ t_{\rm now},\ \phi,\ \omega, \eta$ }
    $H_{t_{\rm now}} \leftarrow  {\tt ApEn}(x[t_{\rm now}-\phi:t_{\rm now}])$ \\
    $E' \leftarrow {\tt CannyEdge1D}\big((H_i)_{i=t_{\rm now}-\eta}^{t_{\rm now}}\big) $ \\
    $E'' \leftarrow {\tt RemoveCloseEdges}(E', \omega) $ \\
    \If{$ t_{\rm now} \in E''$}
    {
      $E \leftarrow E \cup \{ t_{\rm now}\}$
    }
  \end{algorithm}
\end{minipage}

Algorithm~\ref{alg:scd_offline} ({\tt OFLStateTrans}) presents the offline state transition detection. First, it computes {\tt ApEn} for every sub-sequence of length $\phi$. Then, it employs canny edge detection algorithm \cite{cannyedge} on $(H_i)_{i=1}^{|x|-\phi}$, which is an image processing technique for extracting boundaries from images. Since we consider one-dimensional data, {\tt OFLStateTrans} uses a 1D variant of canny edge detection (${\tt CannyEdge1D}$) to analyze the list $(H_i)_{i=1}^{|x|-\phi}$ and identify rapid changes in three steps:

\begin{enumerate}
  \item Smooth the list $(H_i)_{i=1}^{|x|-\phi}$ to remove noise by convolution with a digitalized Gaussian filter.
  \item Take the first derivatives of the smoothed list to obtain the changes. In 1D setting, the peaks in the first derivatives represent potential edges.
  \item Return the edges from the list of potential edges only with large magnitudes.
\end{enumerate}

\begin{figure}[htbp!]
  \centering
  \includegraphics[width=\textwidth]{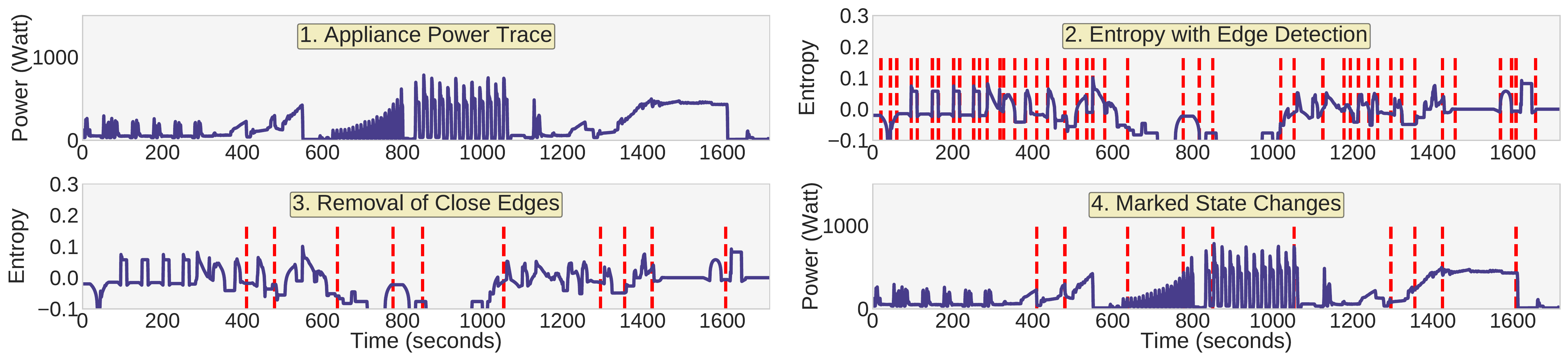}
  \caption{\color{black} An illustration of the state transitions detected by Algorithm~\ref{alg:scd_offline} ({\tt OFLStateTrans}).}
  \Description{An illustration of the state transitions detected by Algorithm~\ref{alg:scd_offline} ({\tt OFLStateTrans}).}
  \label{fig:smartplug:scd_nimd_example}
\end{figure}

Finally, the edges that have less than $\omega$ in separation distance will also be removed. 
{\color{black} 
Fig.~\ref{fig:smartplug:scd_nimd_example} illustrates the state transitions detected by Algorithm~\ref{alg:scd_offline} ({\tt OFLStateTrans}) on a sample trace of power consumption signals. {\tt OFLStateTrans} can correctly detect many state transitions, by which segmentation of the continuous signals can be constructed.
}

\subsubsection{Online Detection}

Algorithm~\ref{alg:scd_online} ({\tt ONLStateTrans}) presents the online state transition detection using only the data observed thus far without complete knowledge of future data. It operates with a sliding window by computing {\tt ApEn} for a window of $\eta$ timeslots. It will decide if an edge is detected at the current time $t_{\rm now}$ based on {\tt CannyEdge1D} using the data of the last  $\eta$ timeslots. The difference between the online and offline algorithms is that the online algorithm is based on myopic information of the last $\eta$ timeslots rather than the full information of every sub-sequences in the whole time horizon. We will evaluate the performance of {\tt ONLStateTrans} in Section~\ref{sec:smartplug:eval}.

\subsection{Classification with Known Classes of Patterns}\label{subsec:smartplug:classif_known_patterns}

Once a state transition is detected, we extract the corresponding segment of signals and classify it by a fitting an analytic model to the signals. There are a number of basic power consumption models suitable for describing the power consumption rates of appliances \cite{appliance_modeling,nilm}:
\begin{enumerate}
  \item  \textit{On-off Model:} This model has a certain fixed power consumption rate $P_{\rm on}$ when active.

  \item  \textit{On-off Decay Model:} The power consumption rate follows an exponential decay curve, dropping from the initial surge power $P_{\rm peak}$ to a stable power $P_{\rm active}$ at a decay rate $\lambda$.
        \begin{equation}
          P(t) = P_{\rm active} + (P_{\rm peak}-P_{\rm active})^{-\lambda t}
        \end{equation}

  \item  \textit{On-off Growth Model:} The power consumption rate follows a logarithmic growth curve, starting with a level $P_{\rm base}$ and a growth rate $\lambda$ .
        \begin{equation}
          P(t) = P_{\rm base} + \lambda \cdot \ln t
        \end{equation}

  \item  \textit{Stable Min-Max Model:} The power consumption rate is characterized by stable power $P_{\rm stable}$ with random upward or downward spikes $P_{\rm spike}$. The magnitude of random spikes is uniformly distributed between $P_{\rm stable}$ and $P_{\rm spike}$, and the inter-arrival times of spikes follow an exponential distribution with mean $\lambda$.

  \item  \textit{Random Range Model:} The power consumption rate is similar to a random walk between a maximum power $P_{\rm max}$ and a minimum power $P_{\rm min}$.

  \item  \textit{Cyclic Model:} The power consumption rate exhibits repetitive patterns.
\end{enumerate}

Resistive and inductive appliances exhibit on-off, on-off decay, or on-off growth behavior, whereas appliances with non-linear power consumption (e.g., with non-sinusoidal current waveform) exhibit stable-min, stable-max, or random-walk behavior. In addition, many composite appliances (e.g., fridge, washing machine, air-conditioner) are composed of a combination of resistive, inductive and non-linear basic loads, which exhibit more complex power consumption patterns.

Given a segment of power consumption signals, the smart plug derives the model that best matches the observed power consumption rate. First, it differentiates the segment and compares its standard deviation with that of the original segment. If the standard deviation has decreased after differentiation, then the segment is best to be fit by a deterministic curve. On the other hand, if the standard deviation has increased, then the segment is best to be fit by a probability distribution.
The details of model fitting are described as follows:

\begin{itemize}
  \item For on-off and on-off growth models, it employs ordinarSection~\ref{sec:smartplug:eval}y linear least squares (OLS) method, which is a mathematical procedure for finding the best fitting curve to a given set of points by minimizing the sum of the squares of the residuals (i.e., the offsets of the points from the curve).

  \item For on-off decay model, it employs a special technique proposed in \cite{exp_fit}. Fitting an exponential decay curve involves three parameters (i.e., $P_{\rm peak}$, $P_{\rm active}$, and $\lambda$). The best fit model is chosen as the one with the least discrepancy.

  \item For stable min, stable max and random range models, we follow the approach proposed in \cite{appliance_modeling} in order to fit a probability distribution to the segment. Specifically, for stable min and stable max models, it derives $P_{\rm spike}$ as the mean of the data plus two standard deviations, and $\lambda$ as the average duration between spikes. To estimate $P_{\rm stable}$, it smooths the data by removing the spikes and estimates $P_{\rm stable}$ from the smoothed data using linear regression. In case of the random range model, the parameters $P_{\rm max}$ and $P_{\rm min}$ are obtained by simply choosing the maximum and minimum values in the data. As before, the best fit model is chosen as the one with the least discrepancy.

  \item For cyclic model, it uses {\em autocorrelation} to determine repeating patterns in the data as proposed in \cite{nilm}. If the data is cyclic, then its autocorrelation will attain a local maximum at each lag that is a multiple of the cycle period. It computes the autocorrelation of the data for up to $n$ lags and sees if the local maxima are separated by similar distances. If so, then the distance between the local maxima is determined as the duration between cycles. %Fig.~\ref{fig:smartplug:cycle} illustrates the cycle detection process using autocorrelation for a portion of the Iron trace in Fig~\ref{fig:smartplug:scd2}.
\end{itemize}

\subsection{Classification of Unknown Patterns}\label{subsec:smartplug:classif_unknown_patterns}

In the previous section, we assumed that a pattern always belongs to a finite number of classes (e.g., power consumption models), which may be known by the smart plug in advance. In this section, we allow these patterns to be unknown to the smart plug.  In this paper, we focus on the approach of online learning of unknown patterns \cite{GGMMc2008cluster}. 
Alternatively, one can learn unknown patterns offline and then using it online. But this requires considerable a-prior training data, more memory and more latency for learning, which is not always desirable in IoT applications. On the other hand, we are able to demonstrate that learning and classifying of unknown patterns in an online manner can be achieved efficiently even using resource-constrained platforms.

For the case of unknown patterns, we employ online clustering algorithm ({\tt ONL$k$MeanCluster}) to group the patterns detected by Algorithm~\ref{alg:scd_online} into clusters such that the patterns within a cluster are similar to each other but different from those in other clusters. 
Each cluster can be regarded as a canonical state of the appliance. We note that the segment length of continuous signals in one cluster may be different since the state transitions do not necessarily occur at fixed time intervals. Thus, we need an abstract representation of the patterns to facilitate consistent clustering. To this end, we use the following five attributes to represent each segment of continuous signal:

\begin{itemize}

  \item Measures of central tendency: (i) \textit{arithmetic mean}, (ii) \textit{median}, and (iii) \textit{mode}.

  \item Measures of statistical dispersion: (iv) \textit{standard deviation}, and (v) \textit{range}.

\end{itemize}

Next, all segments of continuous signals will be clustered based on a tuple of these five attributes.

{\tt ONL$k$MeanCluster} is based on {\em Doubling Algorithm} \cite{Charikar:1997} and {\em Online k-Means Clustering with Discounted Updating Rule} \cite{king2012}, as described in Algorithm~\ref{alg:clust}.  
While there are other clustering algorithms with advanced machine learning capabilities, k-means is often regarded as the simplest approach with efficient implementation and small memory requirement. also, the efficient implementation of K-means clustering algorithm can facilitate the implementations of more sophisticated clustering algorithm with advanced machine learning capabilities.
The algorithm aims to find a suitable solution to the problem given in Eqn.~(\ref{eq:kmin_patterns}). It consists of two stages: {\em the update stage} and {\em the merging stage}. In the update stage, the algorithm adds each segment either to an existing cluster or puts it in a new cluster. This stage continues for as long as the number of clusters is less than or equal to $k$. When the number of clusters exceed $k$, then the algorithm proceeds to the merging stage. In the merging stage, the algorithm reduces the number of clusters by merging clusters that are within a certain distance of each other. The merging stage guarantees that no more than $k$ clusters are selected in the presence of streaming data input. The detailed description of the clustering algorithm is given in Algorithm~\ref{alg:clust}, as follows.

\begin{algorithm}[hbtp!]
  \caption{\tt ONL$k$MeanCluster}
  \label{alg:clust}
  \SetAlgoLined
  \DontPrintSemicolon
  \SetKwInput{Input}{Input}
  \SetKwInOut{Output}{Output}
  \SetKwInOut{Initialization}{Initialization}
  \SetKwComment{Comment}{$\triangleright$\hspace{-4pt}\ }{}\SetInd{0.2em}{0.35em}
  %\SetKwComment{Comment}{$\triangleright$\ }{}
  \Input{Segment at time $t$: $z_t$, Max num. of clusters: $k$, Weight: $\alpha$}
  \Initialization{${\cal C} \gets \big\{\{ z_i\}_{ i =1}^{k}\big\}$, %\hspace{25pt} $\triangleright$ {first k+1 patterns as separate clusters} \\
  ${\cal Z} \gets \left\{z_i\right\}_{ i =1}^{k}$, %\hspace{34pt} $\triangleright$ {each segment itself is the cluster center} \\
  $d^*  \gets \underset{c_1, c_2 \in {\cal Z}:~ c_1 \ne c_2}{\min} \ \| c_1, c_2\|^2_2$ %\hspace{45pt} $\triangleright$ {minimum inter cluster distance in ${\cal C}$}
  }
  %\hrule
  \If(\Comment*[f]{\it Updating clusters}){$|{\cal C}| \le k$}
  {
  $i \gets \underset{i \in \{1,...,k\}}{\arg\min} \ \| z_t, c_i\|^2_2$  \Comment*[r]{\it Get the nearest cluster to $z_t$}
  \eIf{$\| z_t, c_i \|^2_2 \le 2d^*$}
  {
    ${\cal C}_i \gets {\cal C}_i \cup \{z_t\}$   \Comment*[r]{\it Add segment $z_t$ to the nearest cluster ${\cal C}_i $}
    %    $c_i \gets c_i + \frac{1}{|{\cal E}_i|}(z_t - c_i)$   \Comment*[r]{update the cluster center}
    $c_i \gets c_i + \alpha(z_t - c_i)$   \Comment*[r]{\it Update the cluster center}
  }
  {
    ${\cal C} \gets {\cal C} \cup \big\{\{z_t\}\big\}$   \Comment*[r]{\it Create a new cluster comprising of $z_t$}
  }
  }
  \If(\Comment*[f]{\it Merging clusters}){$|{\cal C}| > k$}
  {
  $(i, j)  \gets \underset{i, j \in {\cal Z}\mid i \ne j}{\arg\min} \ \| c_i, c_j\|^2_2$ \Comment*[r]{\it Find two closest clusters}
  ${\cal C}' \gets {\cal C}_i \cup {\cal C}_j$ \Comment*[r]{\it Merge two closest clusters}
  $c' \gets \frac{1}{|{\cal C}'|} \underset{y \in C'}\sum y$ \Comment*[r]{\it Center of the merged cluster}
  ${\cal C} \gets {\cal C} \cup \{{\cal C}' \} \backslash \{{\cal C}_i, {\cal C}_j\}$  \Comment*[r]{\it Update clusters}
  ${\cal Z} \gets {\cal Z} \cup \{c'\} \backslash \{c_i, c_j\}$  \Comment*[r]{\it Update cluster centers}
  $d^* \gets 2d^*$ \Comment*[r]{\it Double the  threshold of creating a new cluster}
  }
\end{algorithm}

\begin{itemize}

  \item {\em Initialization:} The algorithm starts by initializing the first $k$ segments as the initial clusters, where ${\cal C}$ denotes the set of clusters and ${\cal Z}$ denotes the set of cluster centers. Initially, each segment itself is the cluster center since each cluster has only one segment. The minimum inter-cluster distance in ${\cal C}$ is denoted by $d^*$.

  \item  {\em Updating Clusters:} Upon receiving a new segment $z_{t}$, the algorithm finds the cluster whose center is the nearest to $z_t$. If the distance between the nearest cluster center and $z_t$ is less than $2d^*$, then $z_t$ is added to the cluster and the cluster center is shifted proportionally. If, however, the distance is greater than $2d^*$, then a new cluster is created and segment $z_t$ is added to it. Notably, we use the discounted updating rule (i.e., $c_i \gets c_i + \alpha(z_t - c_i)$ ) instead of standard online k-Means updating rule (i.e., $c_i \gets c_i + \frac{1}{|{\cal C}_i|}(z_t - c_i)$) because the discounted update has been shown to provide a better result when the cluster centers are varying over time \cite{king2012}. The weight $\alpha \in (0,1)$ determines the relative weight of the new segment $z_t$, which provides an effect of exponential smoothing.

  \item  {\em Merging Clusters:} Whenever the total number of cluster exceeds $k$, the merging step is invoked to reduce the total number of clusters within $k$. In this step, the algorithm first finds and merges the two closest clusters. Then, it adds the newly merged cluster to ${\cal C}$ and removes both old clusters from ${\cal C}$. Similarly, the center of the new cluster is added to ${\cal Z}$ and the old centers are removed. Finally, the cost of creating new clusters $d^*$ is doubled. The effect of doubling the cost is that, eventually, it will become prohibitively expensive to create new clusters. Thus, the algorithm will be more likely to assign new segments to one of the existing $k$ clusters.

\end{itemize}
We will evaluate the performance of {\tt ONL$k$MeanCluster} in Section~\ref{sec:smartplug:eval}.

%% file: track.tex
\section{Memory-efficient  Occurrence Tracking} \label{sec:smartplug:pattern_track}

In this section, we present the basic ideas, techniques, and algorithms for tracking of the pattern occurrences in the power consumption signals over time intervals of different lengths using the compact memory space of smart plug. In particular, we aim to keep track of the temporal occurrences for each pattern over the following intervals: (1) minute-by-minute occurrences of the past hour, (2) hourly occurrences of the past 24 hours, (3) daily occurrences of the past 30 days, (4) weekly occurrences of the past week, (5) monthly occurrences of the past year, and (6) yearly occurrences of the past year. Such detailed tracking will enable the smart plug to monitor the appliance behavior and usage patterns over a long period as well as identify the correlations in usage patterns. 

However, the difficulty is that the memory space requirements for keeping individual track of every pattern for all of the above time intervals may exceed the small memory space of the smart plug. Thus, we focus on approximate tracking, by obtaining the estimated occurrences of the patterns using a special-purpose data structure called \emph{Count-min Sketch}, which is a space-efficient probabilistic data structure that keeps an approximate count of elements in streaming data \cite{countminsketch}. Count-min sketch grows sub-linearly with the input data by randomly summarizing the data. At any given time, the sketch returns an estimated count of a certain pattern in response to a query. This estimated count is shown to be within a fixed threshold of the ground truth with a certain probability. Count-min sketch is similar to the \emph{Bloom Filter} \cite{bloomfilter}. Bloom filter tells the membership of an element in the set, whereas count-min sketch tracks the approximate counts.

In essence, a count-min sketch employs $K$ hash functions to track the patterns (or items in general) using $M$ counters organized in a 2-dimensional ($K\times\frac{M}{K}$) array referred to as a sketch. The parameter $M$ specifies the memory space occupied by the sketch, whereas $K$ determines the time complexity of the sketch. These parameters are chosen during sketch creation in a way that not only makes them independent of the size of the stream but also bounds the discrepancy of the estimated count of a pattern from the ground truth. Recall that $N_i^{\cal T}$ is the true total number of occurrences of pattern $i \in {\cal C}$ in a certain epoch of time ${\cal T}$ and $\widehat{N}_i^{\cal T}$ is the estimated total number of occurrences. By setting $K = \ln\frac{1}{\delta}$ and $M = \ln \frac{1}{\delta} \cdot \frac{e}{\epsilon}$, we obtain ${\mathbb P}\big(|{\widehat{N}}_i^{\cal T} - N_i^{\cal T}| \ge \epsilon \cdot |{\cal T}|\big) < \delta$, where $\epsilon$ bounds the discrepancy between ${\widehat{N}}_i^{\cal T}$ and $N_i^{\cal T}$, while $\delta$ bounds the probability of discrepancy. This implies that the probability of  discrepancy between the estimated count and the true count of pattern $z$ being more than $\epsilon T$ is at most $\delta$. A proof can be found in \cite{mitzenmacher}.

\subsection{Hash Functions}

To implement count-min sketch, we first need to implement hashing efficiently in the smart plug. Hashing can be regarded as a random projection from a high dimensional space of data to a low dimensional space of hashes. In the count-min sketch, each pattern is mapped by the $K$ independent hash functions, where the $i$-th hash function $h_i$ maps the pattern into one of the $\frac{M}{K}$ counters in the $i$-th row of the sketch. The hash functions are drawn independently at random from a 2-universal class of hash functions in the form of:
\begin{equation} \label{eq:smartplug:hash_functions}
  h(x) = ((ax + b)\ {\rm mod}\ p)\ {\rm mod}\ M.
\end{equation}
The parameters are explained as follows.
\begin{itemize}
  \item $x$ is a numeric representation of the pattern we will hash. Note that this requires a translation function to convert each pattern to a numeric form. To this end, we first calculate the string hash of a pattern using \emph{Secure Hash Algorithm 1 } (SHA1) \cite{sha1}, then convert the string representation of the hash to a 32-bit number. We are constrained to using 32-bit numeric representation instead of 64-bit or 128-bit due to the hardware limitation of the smart plug. Similarly, {SHA1} is chosen due to its ease of implementation in the smart plug.

  \item $p$ is a large prime number so that every possible key $x$ is in the range [0, $p-1$]. To satisfy this constraint, we set $p$ to be the smallest prime number greater than $2^{32}-1$ since any 32-bit positive number must be in the range [0, $2^{32}-1$].

  \item $a$ and $b$ are any numbers in the range [1, $p-1$] and [0, $p-1$], respectively.
\end{itemize}

Every possible combination of $a$ and $b$ in Eqn.~(\ref{eq:smartplug:hash_functions}) results in a different hash function. The total number of possible hash functions is $p(p-1)$ since we have $p$ choices for $b$ and $p-1$ choices for $a$. We can easily generate the $K$ hash functions for count-min sketch, simply by substituting $K$ different combinations of $a$ and $b$ in Eqn.~(\ref{eq:smartplug:hash_functions}). Notably, any hash function drawn from 2-universal family has a property that the probability of two different keys $x_i$ and $x_j$ being mapped into the same counter is $\frac{1}{M}$, where $M$ is total number of counters. In count-min sketch, however, each hash function ranges over a row of the counters instead of all $M$ counters, thus the collision probability is given by \cite{mitzenmacher}:
\begin{equation} \label{eq:smartplug:prob_collision}
  \mathbb{P} \{ h(x_i) = h(x_j) \} = \frac{K}{M}.
\end{equation}

\subsection{Updates on Counters} \label{subsec:smartplug:update_sketch}

There are multiple ways to optimize the performance of count-min sketch. One possible way is to choose proper updating rules on the counters upon each occurrence. For every occurrence $t$ of the $i$-th pattern in the streaming data, we apply each of the $K$ hash functions to obtain $C_{i, h_t^i}$ (the $h_t^i$-th counter in the $i$-th row of the sketch). 
\begin{enumerate}

\item
First, we can update the counter using the following rule:
\begin{equation} \label{eq:smartplug:counter_update1}
  \hspace{-24pt} \text{\bf (Update Rule {\tt U1})} \qquad  C_{i, h_t^i} \gets C_{i, h_t^i} + 1.
\end{equation}

\item
Alternatively, we can perform a more conservative update by using the following rule:
\begin{equation} \label{eq:smartplug:counter_update2}
 \hspace{-24pt} \text{\bf (Update Rule {\tt U2})} \qquad   C_{i, h_t^i} \gets \max\Big\{C_{i, h_t^i}, \underset{j \in \{1, ..., K \}}{\min}\{C_{j, h_t^j}\} + 1\Big\}.
\end{equation}

\end{enumerate}

The modification in Eqn.~(\ref{eq:smartplug:counter_update2}) can achieve improved error performance in practice.  We note that each update operation requires computing only a small number of hash functions and basic arithmetic, which has fast running time.

\iffalse
To illustrate the distinction between the two update rules, an example is given in Fig.~\ref{fig:smartplug:tracking_example}. When a pattern $z$ arrives, we can update the sketch counters using either the regular update in Eqn.~(\ref{eq:smartplug:counter_update1}) or the conservative update in Eqn.~(\ref{eq:smartplug:counter_update2}). For convenience, the counters associated with the pattern $z$ are highlighted in the figure. Fig.~\ref{subfig:smartplug:tracking_example_a} shows the sketch before updating the counters for the pattern $z$.  A regular update simply increments all counters by 1, whereas a conservative update increments only some of the counters, thereby reducing the error that the sketch accumulates with time. The conservative update operation is shown in Fig.~\ref{subfig:smartplug:tracking_example_b}. Before the update, the smallest counter associated with $z$ is 4. Following the conservative update rule, we increment only those counters that are less than 4+1. 
\fi

\subsection{Retrieval from Counters} \label{subsec:smartplug:query_sketch}

Since the count-min sketch randomly summarizes the data in the counters. A proper retrieval rule from the counters will affect the accuracy of approximate tracking. 
\begin{enumerate}

\item
First, we can retrieve the values of counters of the $i$-th pattern by taking the minimum of all the counters mapped by the corresponding hash functions: 
\begin{equation} \label{eq:smartplug:query1}
 \text{\bf (Retrieval Rule {\tt R1})} \qquad    N_i = \underset{i \in \{1,...,K\}}{\min}\{C_{i, h_t^i}\}.
  %N_i = min\{C_{i,h_t^i} : i=1,\dots,k\}
\end{equation}
Note that $N_i$ is an overestimate of the true count of occurrences (i.e., $N_i \geq T_i$), because multiple items can be mapped to the same counter by a hash function. 

\item
On the other hand, to tackle the problem of overestimation, the following alternative rule to Eqn.~(\ref{eq:smartplug:query1}) is proposed in \cite{countmin_improvement2}:
\begin{equation} \label{eq:smartplug:query2}
\text{\bf (Retrieval Rule  {\tt R2})} \qquad    N_i = \underset{i \in \{1,...,K\}}{\rm median} \left\{ C_{i, h_t^i} - \frac{\sum_{j=1}^{K} [C_{i, j} - C_{i, h_t^i}]}{\frac{M}{K} - 1} \right\}.
\end{equation}
Eqn.~(\ref{eq:smartplug:query2}) basically deducts the estimated noise $(\frac{\sum_{j=1}^{K} [C_{i, j} - C_{i, h_t^i}]}{\frac{M}{K} - 1})$ from each of the $K$ counters and returns the median of the resulting $K$ values as the estimated count $N_i$. 

\end{enumerate}

An integrated approach is also possible -- taking the minimum of Eqns.~(\ref{eq:smartplug:query1}) and (\ref{eq:smartplug:query2}) as the estimated count $N_i$, because Eqn.~(\ref{eq:smartplug:query2}) might actually overestimate $N_i$ more than Eqn.~(\ref{eq:smartplug:query1}).

In the experimental evaluation study in Section~\ref{sec:smartplug:eval}, we will evaluate the effectiveness of the update and retrieval rules on practical power consumption data traces in the smart plug.

%% file: testbed.tex
\section{Prototype Implementation}\label{sec:smartplug:testbed}

In this section, we present the hardware prototype implementation for the smart plug. Our smart plug prototype enables extensive testing and evaluations in real-world scenarios with home appliances.

\subsection{Hardware Platform}

The smart plug requires appropriate embedded systems for the timely computation of the classification and tracking tasks. With this in mind, we developed the smart plug prototype based on Arduino-compatible WiFi-enabled \emph{ESP-12s} hardware platform \cite{esp8266}. Unlike other Arduino-family micro-controller boards (e.g., Arduino Uno, Nano, Micro, and Pro), ESP-12s has enough processing power and memory size to compute the required tasks for timely processing. Table~\ref{tab:arduino} summarizes and compares the specifications of ESP-12s with the Arduino-family boards of similar physical footprint.

\begin{table}[htbp!]
  \centering
  \caption{ESP-12s versus the Arduino-family boards.}
  \label{tab:arduino}
  \scalebox{1.0}{
    %\begin{tabular}{@{} c @{} | @{} c @{} | @{} c @{} | @{} c @{} | @{} c @{} | @{} c @{}} \hline \hline
    \begin{tabular}{@{} c @{} | c | c | c | c | c @{}} \hline \hline
      Type & Clock Speed & Static RAM & Flash Memory & WiFi & Cost (as in Jun, 2017) \\ \hline \hline
    ESP-12s & 80 MHz & 82 kB  & 1 MB & yes & $\le$ \$4.0  \\ \hline 
    Arduino Pro/Nano & 16 MHz & 2 kB  & 32 kB & no &  $\le$ \$3.5 \\ \hline
    Arduino Micro & 16 MHz & 2.5kB  & 32 kB & no & $\le$ \$9.0  \\ \hline
%    Raspberry Pi 2 & 900 MHz & 1 GB  & Micro SD & yes & $\le$ \$35.0  \\ \hline 
  \hline
  \end{tabular}}
\end{table}

The list of parts/components used in the smart plug prototype is provided in Table~\ref{tab:append:components}. In our prototype, we have managed to keep the total cost as low as \$20 per device.

\begin{table}[htbp!]
  \centering
  \begin{tabular}{l | c | c} \hline \hline
    Component & Quantity & Cost (as in June 2017) \\ \hline \hline
    ESP-12S Arduino micro-controller (Wemos D1 Mini) & 1 & $\le$ \$4.0 \\ \hline
    Relay (Wemos DC 5V relay shield) & 1 & $\le$ \$2.0 \\ \hline
    Current sensor (ACS712-20A) & 1 & $\le$ \$2.0 \\ \hline
    Analog multiplexer (74HC4052 or 74HC4051) & 1 & $\le$ \$2.0 \\ \hline
    Circuit-mount 220V-5V power supply & 1 & $\le$ \$3.5 \\ \hline
    Capacitors (10 $\mu$F, 440 nF) & 2 & \multirow{2}{*}{ $\le$ \$2.0} \\ \cline{1-2}
    Resistors (100 k$\Omega$, 10 k$\Omega$, 5.1 k$\Omega$) & 5 &  \\  \hline
    Universal size perforated PCB (5 x 7 cm) & 1 & \multirow{3}{*}{ $\le$ \$3.0} \\ \cline{1-2}
    Push button for WiFi & 1 &  \\ \cline{1-2}
    Power jack for 2.1mm for transformer & 1 & \\ \hline \hline
     Total Cost	&  & $\le$ \$20 \\ \hline \hline
  \end{tabular}
  \caption{List of parts/components used in the smart plug prototype .}
  \label{tab:append:components}
\end{table}

\subsection{Smart Plug Prototype}

\begin{figure}[htb!]
  \centering
  \includegraphics[scale=1.2]{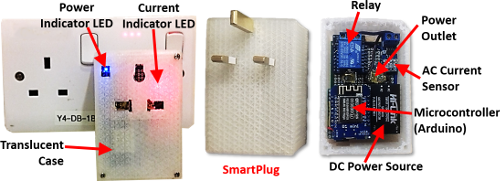}
  \caption{\small Hardware prototype of the smart plug.}
  \Description{\small Hardware prototype of the smart plug.}
  \label{fig:smartplug}
\end{figure}

Fig.~\ref{fig:smartplug} shows the hardware prototype of the smart plug. The smart plug contains a relay to control the power supply to the connected appliance, and several sensors to measure the instantaneous AC voltage and current of the appliance. The AC current is measured using Hall effect-based linear current sensor (ACS712). The AC voltage can be measured using an external 220V--9V step-down voltage transformer. The smart plug can perform the following functions:
\begin{enumerate}

\item {\bf Appliance Monitoring:} The smart plug can perform sophisticated monitoring of the power consumption behavior such as detection, classification, clustering, and tracking of the appliance as described in this paper. These monitoring tasks can enable further advanced functions (e.g., automated demand response, indirect monitoring of the occupant activities, and energy analytics).

\item  {\bf Power Signature Identification:} The ability to measure instantaneous voltage and current enables the computation of a variety of quantities regarding the power consumption, such as active power, reactive power, apparent power, root-mean-square voltage, root-mean-square current, and power factor. The knowledge of these power consumption quantities can give an accurate identification of power consumption signals for appliances, which will be useful for appliance identification, diagnosis, and fault detection.

\item   {\bf Remote Control:} The smart plug supports remote control of the appliance by a WiFi connection. It provides RESTful APIs for controlling the attached appliance and querying its status. The APIs can be accessed by basic HTTP commands from smartphones and web clients. An Internet-connected base station based on Raspberry Pi supports WiFi connections to multiple smart plugs. The smart plugs can transmit the power consumption data and its monitoring results to a third-party base station. The base station provides APIs that can be accessed by smartphones and web clients to visualize the data received from the smart plugs. The base station can provide remote access to the smart plugs for monitoring and control of appliances over the Internet.

\end{enumerate}

\subsection{System Architecture}

To develop the smart plug prototype, we created a \emph{Fritzing} \cite{fritzing} diagram of the breadboard circuit. A Fritzing diagram contains the necessary detail for producing physical circuit boards. It also automatically generates printed circuit boards (PCB) from the circuit design. We created 

\begin{figure*}[htbp!]
  \centering
  \includegraphics[width=0.75\textwidth]{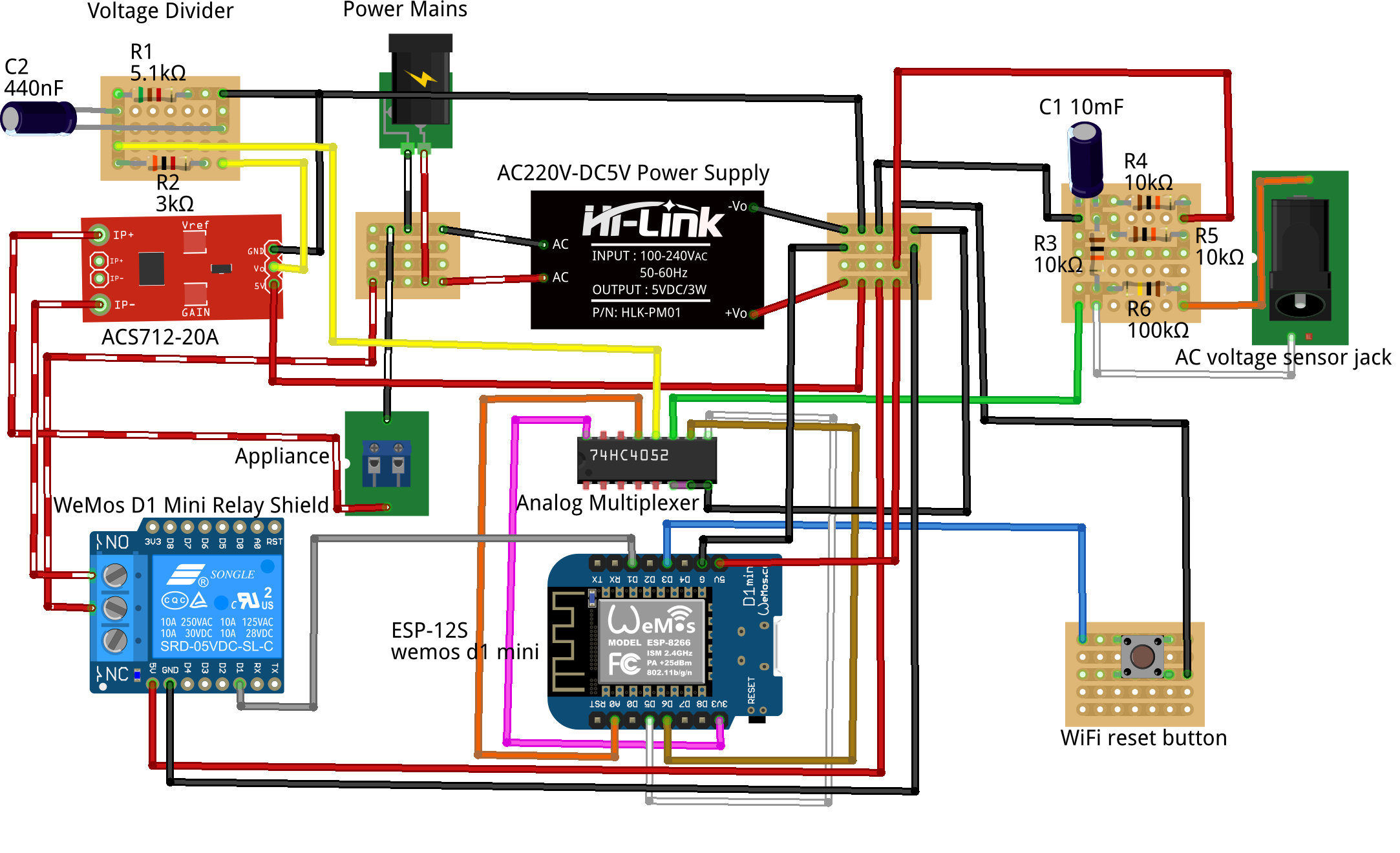}
  \caption{Fritzing diagram of the smart plug prototype.}
  \Description{Fritzing diagram of the smart plug prototype.}
  \label{fig:append:fritzing}
\end{figure*} 

\iffalse
\begin{figure*}[htbp!]
  \centering
  \includegraphics[width=0.75\textwidth]{./figs/smartplug_schem.pdf}
  \caption{Schematic view of the smart plug prototype.}
  \label{fig:append:schematic}
\end{figure*} 
\fi

Fig.~\ref{fig:append:fritzing} depicts the Fritzing diagram for our smart plug prototype, with connectivity to all the components. In particular, the banded red wiring indicates 220V AC power connection, and the banded black wiring is the corresponding neutral connection. Notably, the AC voltage sensor circuit in the upper right corner of Fig.~\ref{fig:append:fritzing} is based on the circuit design in \cite{openenergymonitor_ac_circuit}. An external 220-9V AC-to-AC voltage transformer can be plugged into the AC voltage sensor jack, which will provide instantaneous AC voltage. If the voltage transformer is not connected, then the smart plug uses a fixed RMS voltage to calculate appliance power consumption. The voltage divider in the upper left corner of Fig.~\ref{fig:append:fritzing} is used to level shift the voltage output of the AC current sensor from 5V to 3V. The level shifting is necessary because the analog pin of ESP-12S has an operating voltage range of 0-3.2V. The analog multiplexer in the center of Fig.~\ref{fig:append:fritzing} is also needed because both the AC current sensor and AC voltage transformer provide analog signals, but ESP-12S micro-controller has only one analog input pin. Thus, the multiplexer enables ESP-12S to read both analog inputs. To minimize the form factor of the smart plug, we divided the overall circuit into several layers such that they can be stacked on top of each other.

\subsection{Software System}

%This section includes more software description of the work:  (e.g, what modules / libraries were used, size of code, software architecture etc)

The smart plug is required to carry out timely computation of various tasks involved in event detection, classification, and clustering (e.g., approximate entropy computation, edge detection, smoothing, linear regression, least square fitting, autocorrelation computation, k-means clustering, and hashing, etc.). To implement these tasks, we developed native modules in Arudino (C++) using object-oriented programming paradigm.   

In addition, we also developed a web server along with a client graphical user interface (GUI) that allows for wireless interaction with the smart plug. The server provides RESTful API that can be used to query the power state of the smart plug and request data. The server is also implemented in Arduino (C++) and runs directly on the smart plug. The client GUI can be accessed from a web browser by navigating to the IP address of the smart plug. The smart plug supports two WiFi modes: (i) stand-alone access point (AP) mode, which allows users to directly connect to the smart plug without the need for a WiFi router, (ii) station mode, where the smart plug can connect to an existing WiFi router to become part of the local WiFi network.

Section~\ref{subsec:eval:overhead} provides details regarding the size of the code and the memory requirements of our software system. Moreover, the source code of the smart plug software system is currently released publicly \cite{smartplug_sourcecode} to give other researchers access to all aspects of the implementation and enable them to take advantage of our work for their future projects.

%% file: eval.tex
\section{Experimental Evaluation} \label{sec:smartplug:eval}

In this section, we provide a detailed experimental evaluation study of the techniques and algorithms proposed in this paper and implemented in the smart plug prototype.

\subsection{Pattern Detection} \label{sec:smartplug:eval_pattern}

\begin{figure}[htbp!]
  \begin{subfigure}[t]{\textwidth}
    \centering
    \includegraphics[width=\textwidth]{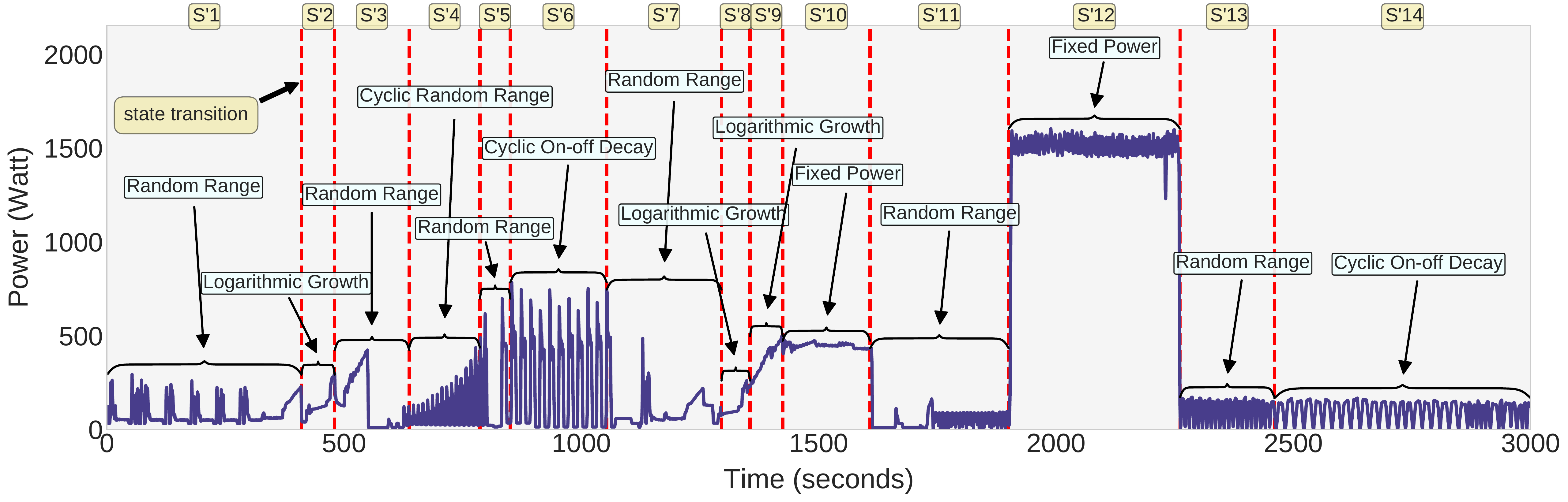}
    \caption{Offline state transition detection using Algorithm~\ref{alg:scd_offline} ({\tt OFLStateTrans}).}
    \Description{Offline state transition detection using Algorithm~\ref{alg:scd_offline} ({\tt OFLStateTrans}).}
    \label{fig:smartplug:scd1a}
  \end{subfigure}
  \begin{subfigure}[t]{\textwidth}
    \centering
    \includegraphics[width=\textwidth]{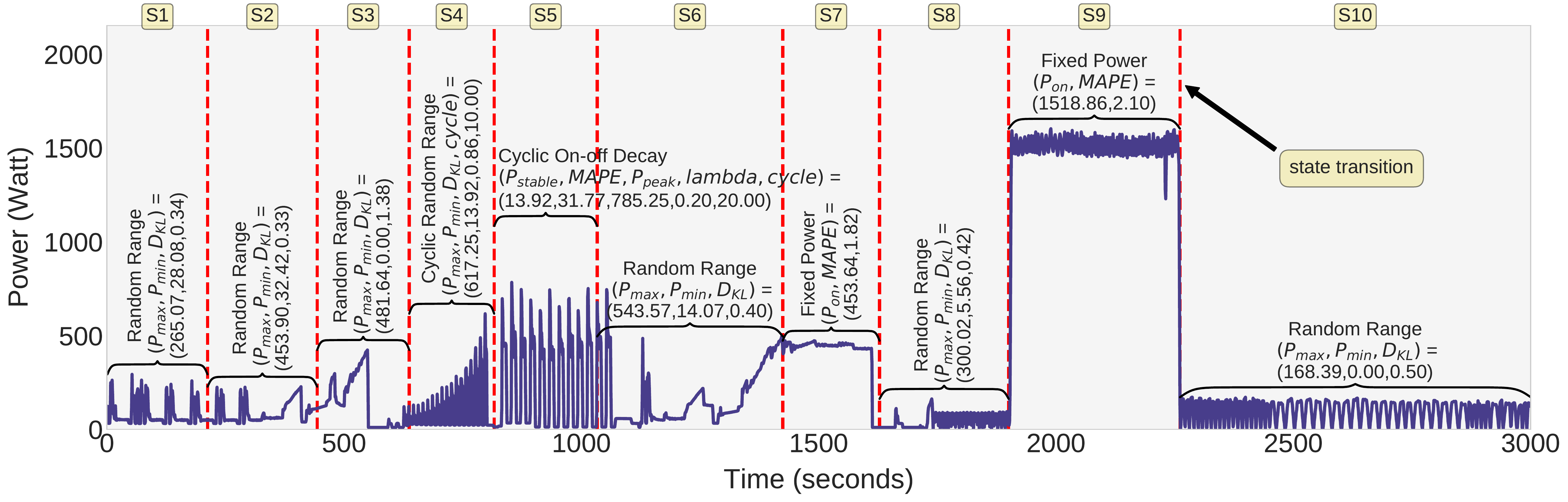}
    \caption{Online state transition detection using Algorithm~\ref{alg:scd_online} ({\tt ONLStateTrans}).}
    \label{fig:smartplug:scd1b}
  \end{subfigure}
  \caption{Results of state transition detection for the power consumption signals of a washing machine.}
  \label{fig:smartplug:scd1}
\end{figure}

\begin{figure}[htbp!]
  \centering
  \includegraphics[width=\textwidth]{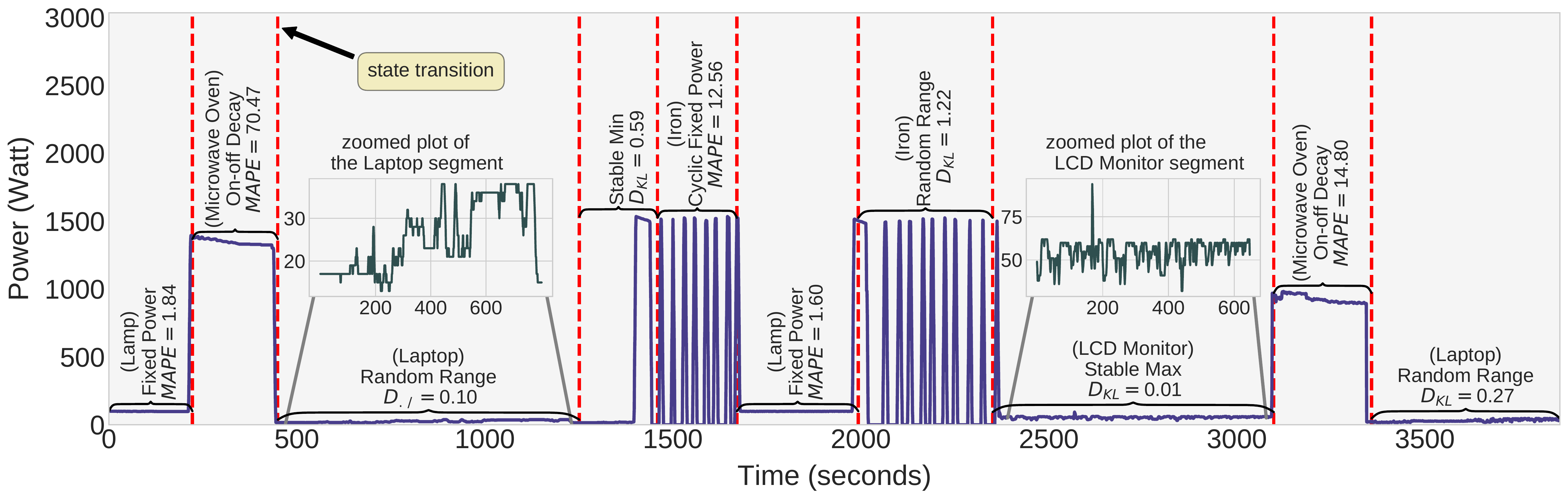}
  \caption{Results of online state transition detection using Algorithm~\ref{alg:scd_online} ({\tt ONLStateTrans}) for the power consumption signals with multiple appliances.}
  \Description{Results of online state transition detection using Algorithm~\ref{alg:scd_online} ({\tt ONLStateTrans}) for the power consumption signals with multiple appliances.}
  \label{fig:smartplug:scd2}
\end{figure}

In this section, we discuss the performance evaluations of both offline and online algorithms. The results of state transition detection for a sample trace of power consumption of a washing machine are illustrated in Fig.~\ref{fig:smartplug:scd1}. The washing machine is a composite load with several active power states, each corresponding to a constituent basic load. Each vertical dashed line in Fig.~\ref{fig:smartplug:scd1} indicates a state transition event. In particular, Fig.~\ref{fig:smartplug:scd1a} shows the results by Algorithm~\ref{alg:scd_offline} ({\tt OFLStateTrans}), which can be compared with the results of  Algorithm~\ref{alg:scd_online} ({\tt ONLStateTrans})  in Fig.~\ref{fig:smartplug:scd1b}. The parameter values used by the algorithms during the experiment are listed in Table~\ref{tab:smartplug:scd_params}. { The table shows that both algorithms use sliding window (of length $\phi$) in order to compute {\tt ApEn}. However, Algorithm~\ref{alg:scd_offline} operates over a sliding lookahead window assuming all future data is available, whereas Algorithm~\ref{alg:scd_online} operates over a sliding lookback window of a limited period of data.} We observe that both algorithms are able to partition the sample trace into segments of states, as indicated by the vertical dashed lines in Figure~\ref{fig:smartplug:scd1}. However, there are some differences between the results of the two algorithms in terms of the number of detected states and the respective times. In particular, {\tt OFLStateTrans} has detected 14 states (S'1--S'14), while {\tt ONLStateTrans} has returned only 10 states (S1--S10). { The reason of this discrepancy is the unavailability of future data, which makes it hard for online algorithm to detect transitions between similar states.}
\begin{table}[htbp!]
  \small
  \centering \setlength\tabcolsep{3pt}
  \begin{tabular}{c | c | c | c | c | c } \hline \hline
                                    & Edge                 & Sliding         & Sub-sequence & Similarity                                          & Online        \\
                                    & Separation           & Window          & Length       & Threshold                                           & Window        \\
                                    & Threshold ($\omega$) & Length ($\phi$) & ($m$)        & ($\theta$)                                          & Size ($\eta$) \\\hline\hline
    Algorithm~\ref{alg:scd_offline} & 60                   & 60              & $\phi/4$     & $0.2 \cdot \sigma(x[1:N])$                          & -             \\ \hline
    Algorithm~\ref{alg:scd_online}  & 200                  & 60              & $\phi/4$     & $0.2 \cdot \sigma(x[t_{\rm now}-\phi:t_{\rm now}])$ & 7             \\ \hline \hline
  \end{tabular}
  \caption{Values of parameters in the algorithms. $\sigma(\cdot)$ is the standard deviation of a time series.}
  \label{tab:smartplug:scd_params}
\end{table}
We also evaluated the results of state transition detection for a sample trace with multiple appliances in Fig.~\ref{fig:smartplug:scd2}. The trace is comprised of the power data from multiple appliances (i.e., lamp, microwave oven, laptop computer, LCD monitor, and iron), where at any given time only one appliance is connected to the smart plug.
{ We intend to show that our system can automatically classify and track the consumption patterns, even when different appliances are plugged in at different times. Note that our smart plug is not trying to disaggregate the power data with multiple appliances at the same time.}
% Appliances are often plugged into and out of outlets, especially in shared spaces such as living rooms or kitchens.
It can be observed in Fig.~\ref{fig:smartplug:scd2} that both the microwave oven and the iron have a high power consumption rate of more than 1000 Watts, whereas the laptop computer and the LCD monitor consume less than 100 Watts. Due to this large difference between their power consumption levels, it is difficult to depict the precise power consumption variations of laptop computer and LCD monitor from Fig.~\ref{fig:smartplug:scd2}. Thus,  portions of the sample trace corresponding to the laptop computer and the LCD monitor are zoomed in and shown using small plots inside Fig.~\ref{fig:smartplug:scd2}. Again, {\tt ONLStateTrans} is able to partition the sample trace into segments corresponding to the appliances through detection of state transitions from one appliance to another.

{
\begin{figure}[htbp!]
  \begin{subfigure}[t]{0.49\textwidth}
    \centering
    \includegraphics[width=\textwidth]{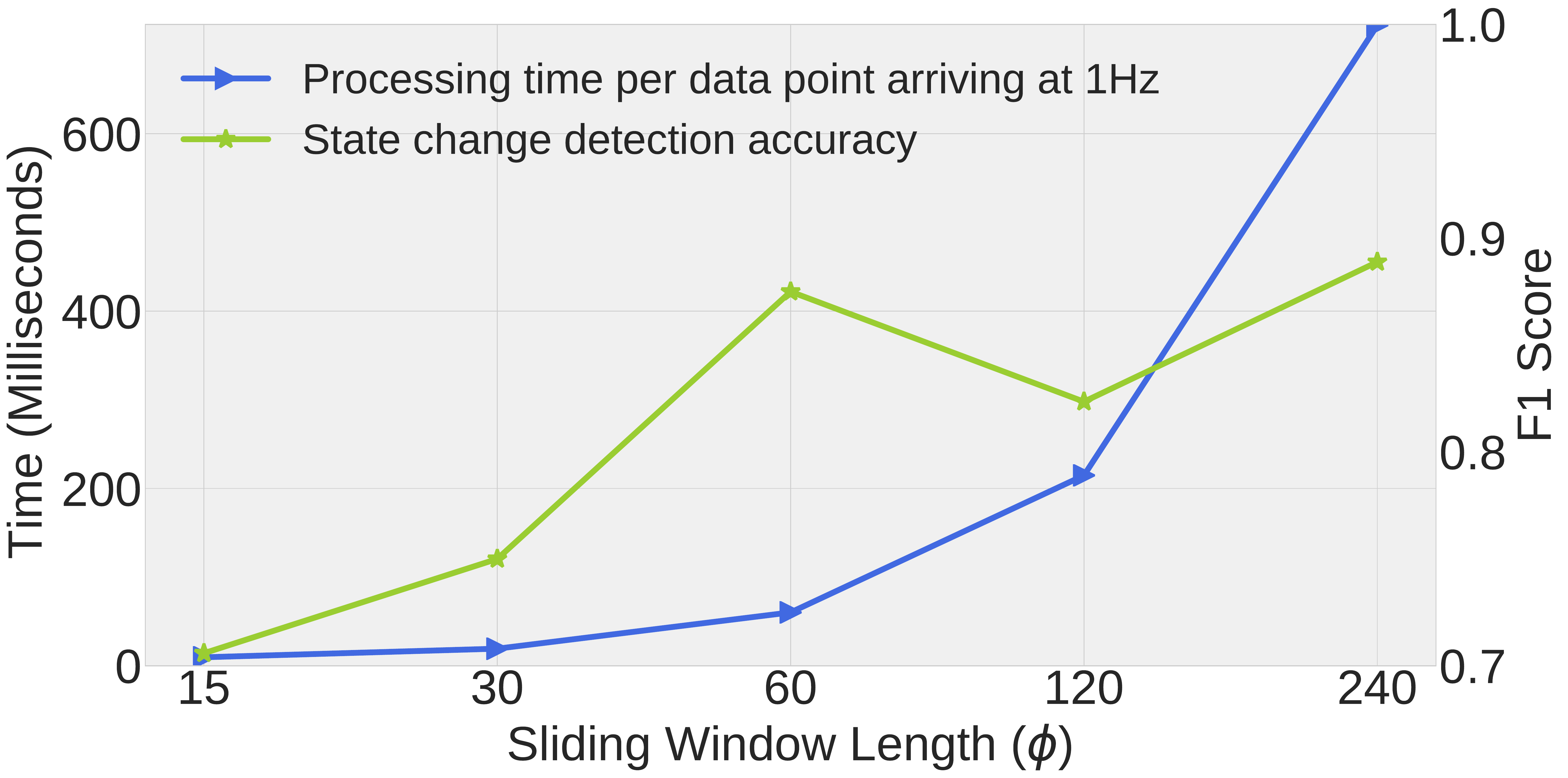}
    \caption{ Effect of the sliding window length.}
    \Description{Effect of the sliding window length.}
    \label{fig:smartplug:scd_eval_a}
  \end{subfigure}
  \begin{subfigure}[t]{0.49\textwidth}
    \centering
    \includegraphics[width=\textwidth]{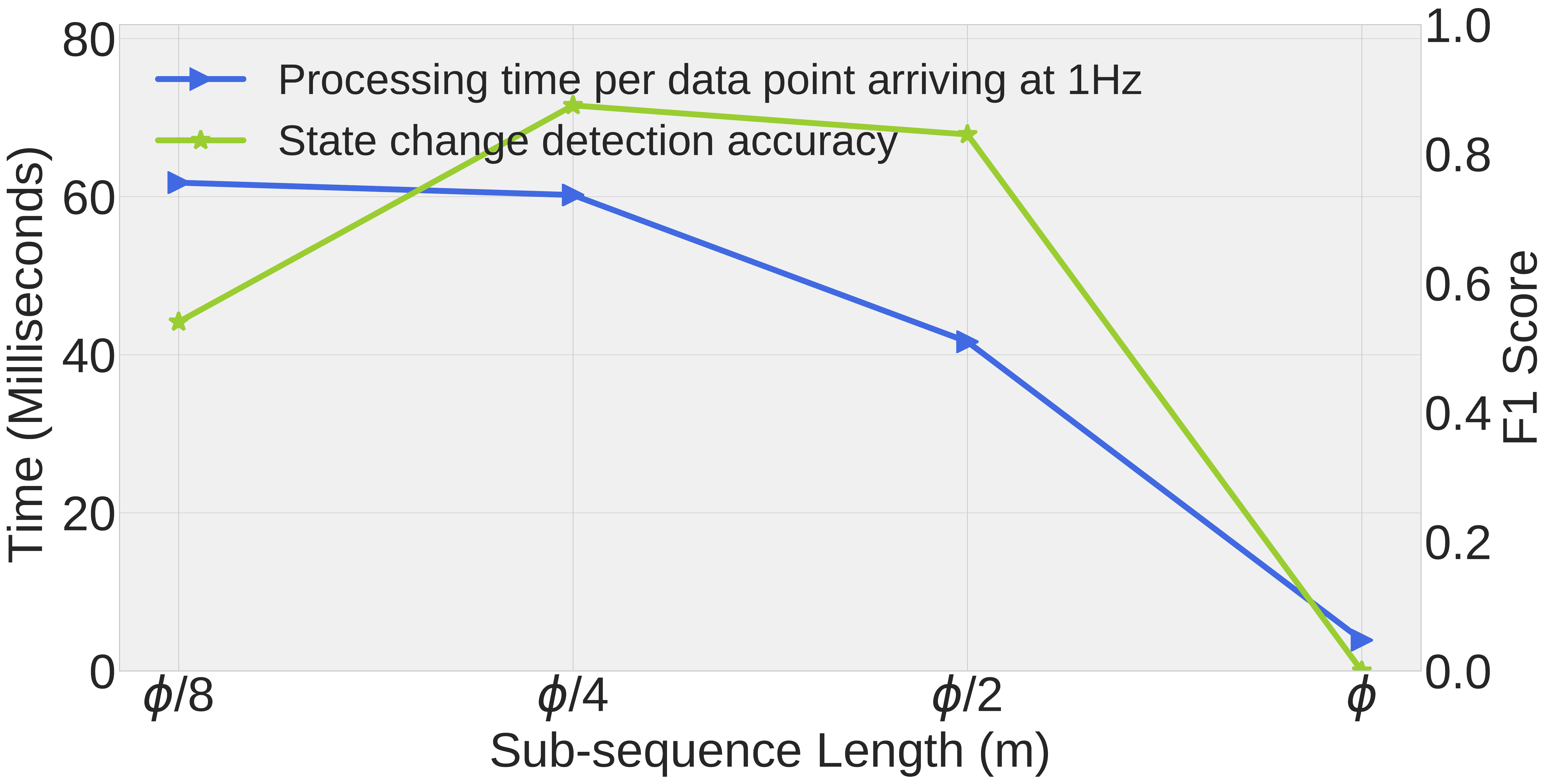}
    \caption{ Effect of the sub-sequence length.}
    \Description{Effect of the sub-sequence length.}
    \label{fig:smartplug:scd_eval_b}
  \end{subfigure}
  \caption{ Trade-off between performance and scalability of Algorithm~\ref{alg:scd_online} ({\tt ONLStateTrans}).}
  \label{fig:smartplug:scd_eval}
\end{figure}

The trade-off between performance and scalability is important for resource-constrained IoT devices like our smart plug. Therefore, we evaluated the smart plug with different values of the parameters listed in Table~\ref{tab:smartplug:scd_params} to demonstrate this trade-off. In particular, we run the experiment with sliding windows and sub-sequences of different lengths and present the results in Fig.~\ref{fig:smartplug:scd_eval}. The primary y-axis represents the processing time per data point arriving at 1 Hz (every second) and secondary y-axis show the performance of the algorithm in terms of F1 score. In Fig.~\ref{fig:smartplug:scd_eval_a}, we can see that as we increase $\phi$, the performance is improved but the time complexity also increases. However, the performance gain beyond $\phi$=60 is negligible while the increase in time complexity is more than linear. Therefore, we use $\phi$=60 in our smart plug platform. Similarly, in Fig.~\ref{fig:smartplug:scd_eval_b}, we can see that the algorithm achieves the best performance (highest F1 score) when m=$\frac{\phi}{4}$. The F1 score worsens when m=$\frac{\phi}{2}$ and drops to 0 when m=$\phi$. The proper values for the remaining parameters in Table~\ref{tab:smartplug:scd_params} were calculated in a similar fashion.
}

\subsection{Classification with Known Classes of Patterns}  \label{sec:smartplug:eval_knownpattern}

The results for classification according to known classes of patterns are depicted in Fig.~\ref{fig:smartplug:scd1}. In particular, Fig.~\ref{fig:smartplug:scd1a} depicts the learned models for segments, whereas Fig.~\ref{fig:smartplug:scd1b} depicts the learned models along with the inferred parameters.

First, we discuss the performance of deterministic curve fitting. Originally,  \cite{nilm} proposes to employ non-linear least square method for curve fitting. Here, we instead employ linear least square method because non-linear least square method involves iterative optimization and requires more memory space that can be provided in the smart plug. Generally, both approaches are able to identify the best fit models of power consumption rates with reasonable accuracy. Furthermore, we observe that the classification outcomes of both approaches are comparable. For instance, the same model has been detected by both approaches for the following pairs of segments in in Figs.~\ref{fig:smartplug:scd1a} and \ref{fig:smartplug:scd1b}: (i) $(S'3,S3)$, (ii) $(S'6, S5)$, and (iii) $(S'12, S9)$.
In Fig.~\ref{fig:smartplug:scd2}, we observe that with the exception of iron, our approach can identify  the same model if it sees such an appliance again.

We evaluate the discrepancy of deterministic curve fitting by Mean Absolute Percentage Error ({\tt MAPE}), defined as follows:
\begin{equation}\label{eq:smartplug:mape}
  {\tt MAPE} = \frac{100}{N} \sum_{t=1}^{N} \frac{P_t - \hat{P_t}}{P_t}
\end{equation}
where $P_t$ is the actual power consumption at time $t$, $\hat{P_t}$ is the predicted power consumption at time $t$, and $N$ is the total number of data points in the segment. A lower value of {\tt MAPE} indicates better accuracy. We provide the values of {\tt MAPE} in Fig.~\ref{fig:smartplug:scd1b} and Fig.~\ref{fig:smartplug:scd2} in the respective segments { (i.e., segments with curve fitting)}, which have reasonably low {\tt {\tt MAPE}} values in all cases except microwave oven, because the particular segment contains slightly imprecise detection of the state transition by {\tt ONLStateTrans}.

Next, we evaluate the discrepancy of probability distribution fitting by Kullback-Leibler (KL) Divergence \cite{kl_divergence}, defined as follows:
\begin{equation}\label{eq:smartplug:d_kl}
  D_{\rm KL}(Y||X) = \sum_{i=1}^{N} Y(i) \log\left(\frac{Y(i)}{X(i)}\right)
\end{equation}
where $Y$ is the true probability distribution of the segment and $X$ is the learned probability distribution. KL divergence indicates the information loss incurred by fitting a specific probability distribution to the data. We provide the values of KL divergence in Fig.~\ref{fig:smartplug:scd1b} and Fig.~\ref{fig:smartplug:scd2} in the respective segments { (i.e., segments with distribution fitting)}. A lower value of KL divergence indicates better accuracy. Based on the $D_{\rm KL}$ values in Fig.~\ref{fig:smartplug:scd1b} and Fig.~\ref{fig:smartplug:scd2}, we observe that the learned probability distribution is a good approximation to the true probability distribution in each segment.

\subsection{Classification of Unknown Patterns}  \label{sec:smartplug:eval_unknownpattern}

\begin{figure*}[!htbp]
  \centering
  \includegraphics[width=\textwidth]{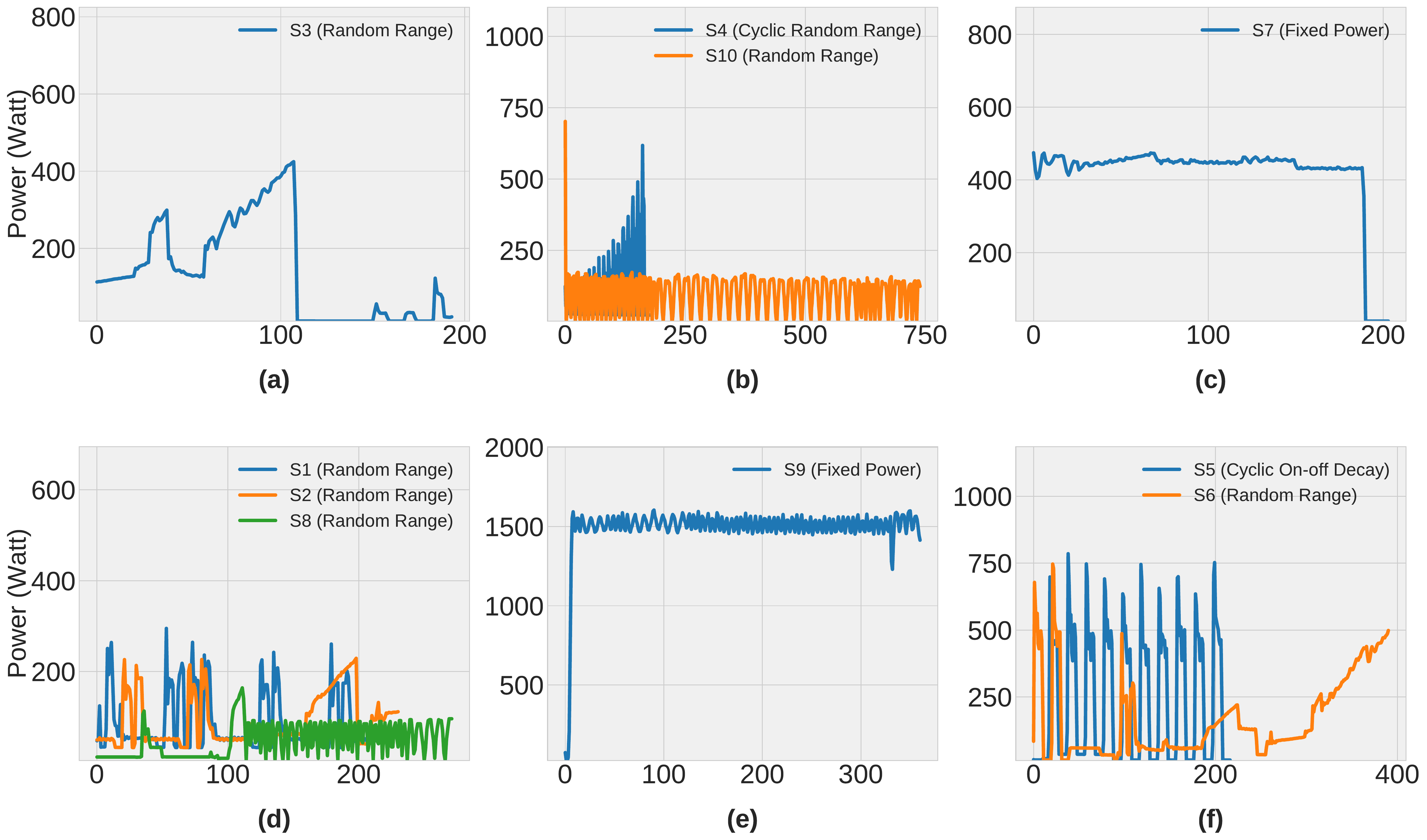}
  \caption{Results of classification of unknown patterns using Algorithm~\ref{alg:clust} ({\tt ONL$k$MeanCluster}). Each sub-figure represents a cluster. X-axis indicates time in seconds for the signals. { The legends indicate the segment ID along with their learned models in Fig.~\ref{fig:smartplug:scd1b}.}}
  \Description{Results of classification of unknown patterns using Algorithm~\ref{alg:clust} ({\tt ONL$k$MeanCluster}). Each sub-figure represents a cluster. X-axis indicates time in seconds for the signals. { The legends indicate the segment ID along with their learned models in Fig.~\ref{fig:smartplug:scd1b}.}}
  \label{fig:smartplug:clust}
\end{figure*}

Fig.~\ref{fig:smartplug:clust} depicts the results of Algorithm~\ref{alg:clust} ({\tt ONL$k$MeanCluster}) when applied to cluster the segments in Fig.~\ref{fig:smartplug:scd1}. 
%The number of clusters $k$ and $\alpha$ were set to 6 and 1.0, respectively. 
There are 6 sub-figures in total, each representing a cluster. The plotted curves in the sub-figures represent segments of power consumption signals that are clustered together by {\tt ONL$k$MeanCluster}. It can be observed that similar segments are indeed clustered together by {\tt ONL$k$MeanCluster}. For instance, all three segments in Fig.~\ref{fig:smartplug:clust}d behave like a random range model. Likewise, the segments in Fig.~\ref{fig:smartplug:clust}b and Fig.~\ref{fig:smartplug:clust}f have cyclic patterns.

To evaluate the results of clustering, we use Sum of Squared Error ({\tt SSE}), which is a standard metric used to measure the goodness of clustering without reference to external information. In particular, we use Within-cluster Sum of Squares Error ({\tt WSSE}) and Between-clusters Sum of Squares Error ({\tt BSSE}), computed by the following equations:
\begin{equation}
  {\tt WSSE}  = \sum_{i}^{k}\sum_{z \in {\cal C}_i} (z - c_i)^2, \qquad {\tt BSSE}  = \sum_{i}^{k}|{\cal C}_i| (c - c_i)^2
\end{equation}
where $z$ is a pattern, ${\cal C}_i$ is the $i$-th cluster, $c_i$ is the center of the $i$-th cluster, $|{\cal C}_i|$ is the size of cluster $i$, and $c$ is the center of all clusters. {\tt WSSE} measures how closely related are the patterns in a cluster (i.e., cohesion), while {\tt BSSE} measures how well-separated the clusters are from each other (i.e., separation).
A good value for $k$ is the one that minimizes ${\tt CH} = \frac{{\tt BSSE}/(k-1)}{{\tt WSSE}/(n-k)}$, where $n$ is total number of patterns, and {\tt CH} is Calinski-Harabasz Index \cite{chk}. Table~\ref{tab:smartplug:clust_error} lists the {\tt WSSE}, {\tt BSSE}, and {\tt CH} for different values of $k$ when clustering the segments in Fig.~\ref{fig:smartplug:scd1} using {\tt ONL$k$MeanCluster}.

\begin{table}[htbp!]
  \centering
  \scalebox{0.75}{
    \setlength\tabcolsep{1pt} % default value: 6pt
    \begin{tabular}{p{2.8cm} | c | c | c | c | c | c | c } \hline \hline
      Num. of clusters ($k$)             & 2                  & 3                 & 4                   & 5              & 6                 & 7                 & 8                 \\\hline
      Output clusters                    & (S1,...,S8,S10);   & (S5,S7,S10);      & (S9); (S7);         & (S1,S2); (S3); & (S3); (S7); (S9); & (S3); (S1,S2,S8); & (S6,S10); (S5);   \\
                                         & (S9)               & (S1,S2,S3,S4,S8); & (S1,S2,S3,S8);      & (S7); (S9);    & (S1,S2,S8);       & (S4,S10); (S5);   & (S4); (S3); (S7); \\
                                         &                    & (S9)              & (S4,S5,S6,S10)      & (S4,S5,S6,S10)
                                         & (S4,S10); (S5,S6); & (S6); (S7); (S9)  & (S8); (S9); (S1,S2)                                                                              \\\hline
      %Cohesion ({\tt WSSE})              & 16133606           & 16897990          & 17659380            & 18420765       & 19182151          & 19943536          & 20704921          \\\hline
      Cohesion ({\tt WSSE})              & 16991017           & 18201188          & 19411360            & 20621532       & 21831704          & 23041875          & 24252047          \\\hline
      %Separation ({\tt BSSE})            & 50508433           & 67941480          & 85362412            & 102783313      & 120204213         & 137625114         & 155046014         \\\hline
      Separation ({\tt BSSE})            & 49237745           & 66216277          & 83194810            & 100173343      & 117151876         & 134130409         & 151108941         \\\hline
      %Calinski-Harabasz Index ({\tt CH}) & 25.04              & 14.07             & 9.66                & 6.97           & 5.01              & 3.45              & 2.13              \\\hline\hline
      Calinski-Harabasz Index ({\tt CH}) & 23.1               & 12.7              & 8.5                 & 6.0            & 4.2               & 2.9               & 1.7               \\\hline\hline
    \end{tabular}}
  \caption{Performance evaluation of Algorithm~\ref{alg:clust} ({\tt ONL$k$MeanCluster}).}
  \label{tab:smartplug:clust_error}
\end{table}

The second row in the table lists the resulting clustering of the segments for the given value of $k$, where each tuple represents a separate cluster. The last row shows that the clustering performance is improved as $k$ increases, but this also increases the number of clusters and the memory space.

\begin{figure}[!htbp]
  \centering
  \includegraphics[width=0.5\textwidth]{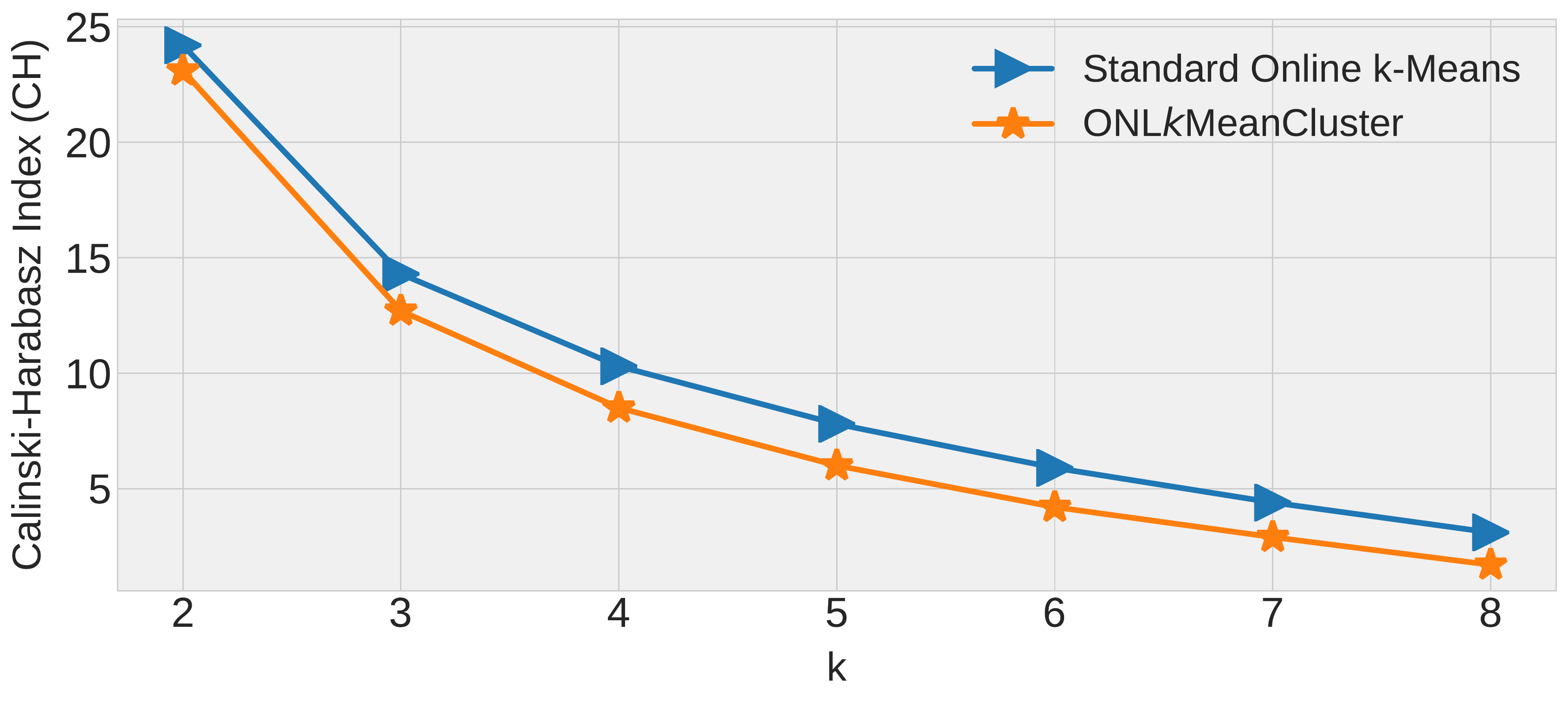}
  \caption{ Comparison between {\tt ONL$k$MeanCluster} and standard $k$-means.}
  \Description{Comparison between {\tt ONL$k$MeanCluster} and standard $k$-means.}
  \label{fig:smartplug:clust_comparison}
\end{figure}

{
Our results show that the results of clustering are acceptable when $k$=6 because similar segments are clustered together. Also, as discussed in Section~\ref{subsec:smartplug:classif_known_patterns}, there are six basic power consumption models that describe the power consumption rates of most appliances. Therefore, choosing $k$=6 will ideally result in a separate cluster for each model and is, therefore, a logical choice.

A general problem with the online clustering algorithms is that the clustering quality may decrease when data streams evolve over time, causing the cluster centers to also shift over time.
To address this problem, Algorithm~\ref{alg:clust}, uses the discounted update rule which has been shown to yield comparatively better results when the cluster centers are evolving over time \cite{king2012}.
In particular, we choose $\alpha$=1.0 in the discounted update rule, which creates the effect of exponential smoothing such that Algorithm~\ref{alg:clust} forgets the initial cluster centers over time and performs close to the optimal solution.
%Finally, {\em k-Means} is sensitive to the initial choice of cluster centers. To remedy this, Algorithm~\ref{alg:clust} uses $\alpha$ which creates the effect of exponential smoothing.   
Another typical problem with the standard {\em $k$-means} is that it may not be a good choice of the algorithm if the number of clusters is not known a priori. However, in the case of Algorithm~\ref{alg:clust}, we already know the number of clusters in advance (i.e., $k$=6).

Figure~\ref{fig:smartplug:clust_comparison} compares the performance of Algorithm~\ref{alg:clust} against the standard online $k$-means. As explained in Section~\ref{subsec:smartplug:classif_unknown_patterns}, Algorithm~\ref{alg:clust} uses the discounted updating rule (i.e., $c_i \gets c_i + \alpha(z_t - c_i)$ ) instead of the standard online $k$-means updating rule (i.e., $c_i \gets c_i + \frac{1}{|{\cal C}_i|}(z_t - c_i)$). It can be seen that Algorithm~\ref{alg:clust} outperforms standard online $k$-means for all values of $k$.
}

\subsection{Occurrence Tracking} \label{subsec:eval_occurrence_tracking}
%{\color{red} \textbf{TODO:} test the proposed scheme with different resource (memory space) constraints because it can help to tell whether it is flexible to be applied to various scenarios. Discuss the implications of relaxing/restricting memory constraints on the performance (reviewer 3)} 
In this section, we present the evaluation results of occurrence tracking in the smart plug prototype. We tested various combinations of the update and retrieval rules from Sections~\ref{subsec:smartplug:update_sketch} and ~\ref{subsec:smartplug:query_sketch}. More specifically, we obtain evaluation results for occurrence tracking using the count-min sketch for the following combinations of update and retrieval rules:
\begin{enumerate}

  \item {\em Count-min} ({\tt CM}): Count-min sketch with regular update rule (Eqn.~(\ref{eq:smartplug:counter_update1})) and regular retrieval rule (Eqn.~(\ref{eq:smartplug:query1})).

  \item {\em Count-min with conservative update} ({\tt CM-CU}): Count-min sketch with conservative update rule (Eqn.~(\ref{eq:smartplug:counter_update2}) and regular retrieval rule (Eqn.~(\ref{eq:smartplug:query1})).

  \item {\em Count-mean-min} ({\tt CMM}): Count-min sketch with regular update rule (Eqn.~(\ref{eq:smartplug:counter_update1})) and a retrieval rule which takes the minimum of Eqns.~(\ref{eq:smartplug:query1}) and ~(\ref{eq:smartplug:query2}).

  \item {\em Count-mean-min with conservative update} ({\tt CMM-CU}): Count-min sketch with conservative update (Eqn.~(\ref{eq:smartplug:counter_update2})) and the same retrieval rule as {\tt CMM}.
\end{enumerate}

Table~\ref{tab:smartplug:track} summarizes the above settings, indicating the update and retrieval rules along with the strength and weakness for each sketch. For example, {\tt CMM-CU} uses update rule (Eqn.~(\ref{eq:smartplug:counter_update2})), which reduces the overestimation during the update operations compared to update rule (Eqn.~(\ref{eq:smartplug:counter_update1})). %Similarly, it uses retrieval rule $min\{1,2\}$, which removes the estimated noise during retrieval operations.

\begin{table}[htbp!]
  \centering
  \scalebox{0.9}{
    \begin{tabular}{l | c | c | m{4cm} | l}\hline\hline
      Setting      & Update Rule & Retrieval Rule           & Strength                              & Weakness         \\ \hline \hline
      {\tt CM}     & {\tt U1}    & {\tt R1}                 & Simplicity                            & Overestimation   \\ \hline
      {\tt CM-CU}  & {\tt U2}    & {\tt R1}                 & Reduced overestimation                & More computation \\ \hline
      {\tt CMM}    & {\tt U1}    & min\{{\tt R1},{\tt R2}\} & Reduced noise                         & More computation \\ \hline
      {\tt CMM-CU} & {\tt U2}    & min\{{\tt R1},{\tt R2}\} & Reduced noise, reduced overestimation & More computation \\ \hline \hline
    \end{tabular}}
  \caption{Settings of different count-min sketches used in our tests.}
  \label{tab:smartplug:track}
\end{table}

For each setting, an evaluation study is conducted where the arrival patterns are first tracked by the minute-by-minute sketch. Then, the minute-by-minute patterns are aggregated into an hourly sketch. Next, the hourly patterns are aggregated into a daily sketch. { The monthly and yearly sketches are computed in a similar manner. This way, we use different sketches (and therefore counters) to store occurrences at each level. At every level of the occurrence tracking (e.g., daily, monthly), the previous level sketch is reset after aggregation of its occurrence counts. For example, the counters in the hourly sketch are reset to zero after the daily occurrences are counted and stored in a daily sketch.}
The patterns are generated by our pattern detection and classification algorithms from synthetic streaming data of power consumption signal { from a single appliance}, comprising of more than one year's data. To compare the errors between the estimated and the actual counts, we compute the exact counts of the patterns in every sketch (i.e., hourly, daily, and monthly). After processing all patterns in the streaming data, we query the sketches to generate approximate occurrence counts of the patterns. { We set $M = 500$ and $K = 5$ during the experiments. Increasing $M$ and $K$ will improve tracking accuracy by reducing over-estimation. However, the smart plug has limited memory capacity, which means we cannot increase the parameter values indefinitely.}

To measure the accuracy of the sketches, we use Mean Absolute Error ({\tt MAE}) and Mean Relative Error ({\tt MRE}). {\tt MAE} is defined as the average of the absolute differences between the estimated and the actual counts of the patterns in a sketch, whereas {\tt MRE} is obtained by dividing {\tt MAE} by the total number of true occurrences of all patterns in the sketch:
\begin{equation} \label{eq:smartplug:error}
  {\tt MAE}  = \frac{\sum_{i=1}^{n} |N(i) - T(i)|}{n}, \qquad
  {\tt MRE} = \frac{\sum_{i=1}^{n} |N(i) - T(i)|}{nT}
\end{equation}

{\tt MAE} and {\tt MRE} provide different insights. {\tt MAE} measures the average magnitude of the errors and provides an indication of how big of an error we can expect from the count-min sketch on average. {\tt MRE}, on the other hand, indicates  how good the tracking performance of a sketch is relative to the number of the total patterns being tracked by the sketch. Note that the {\tt MRE} of one sketch might be considerably smaller than that of another sketch, even though both sketches might have the same value of {\tt MAE}.

\begin{figure}[htbp!]
  \centering
  \includegraphics[width=\textwidth]{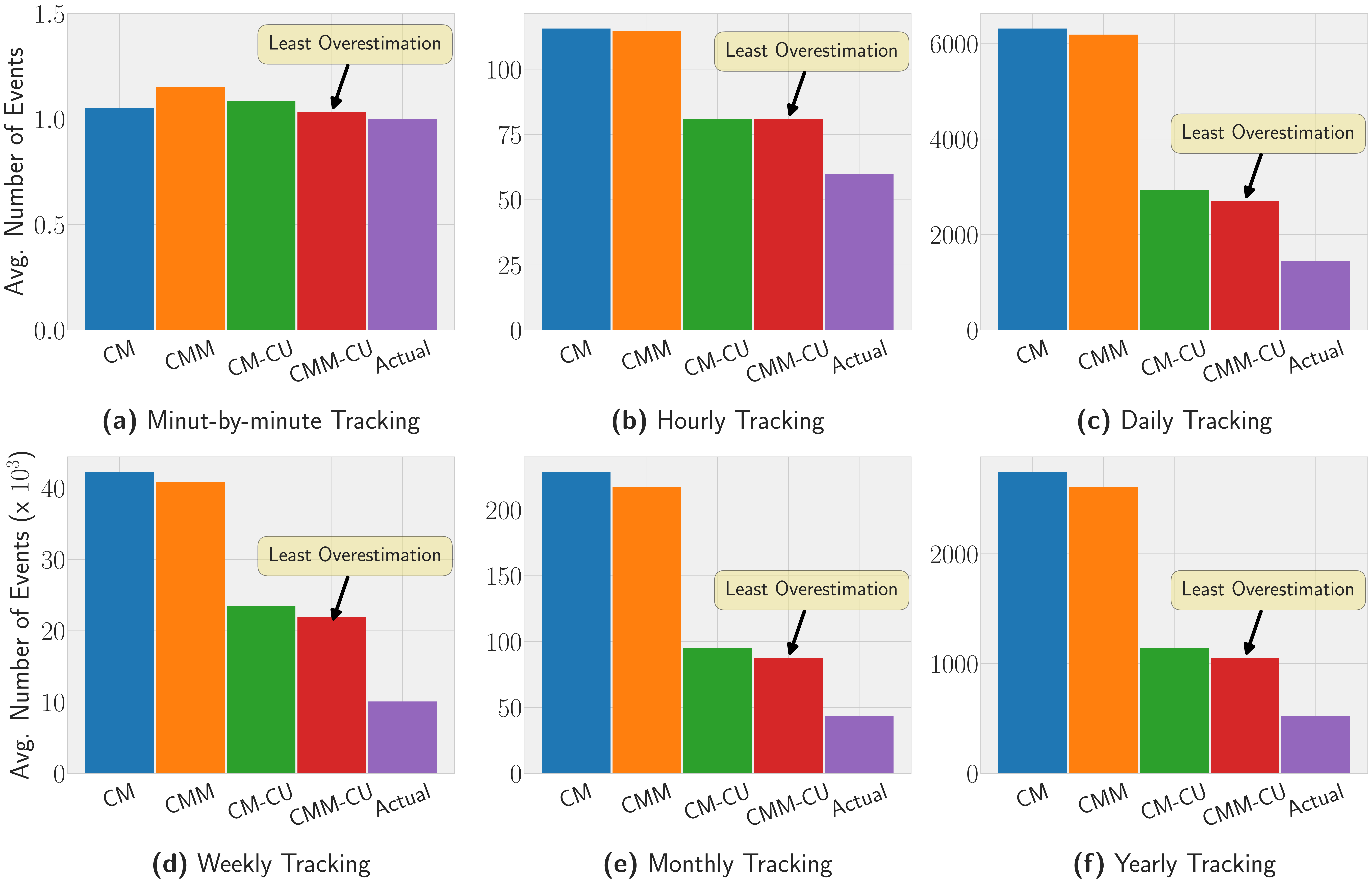}
  \caption{Results of occurrence tracking using count-min sketch.}
  \Description{Results of occurrence tracking using count-min sketch.}
  \label{fig:smartplug:pattern_tracking}
\end{figure}

\begin{figure}[htbp!]
  \centering
  \includegraphics[width=\textwidth]{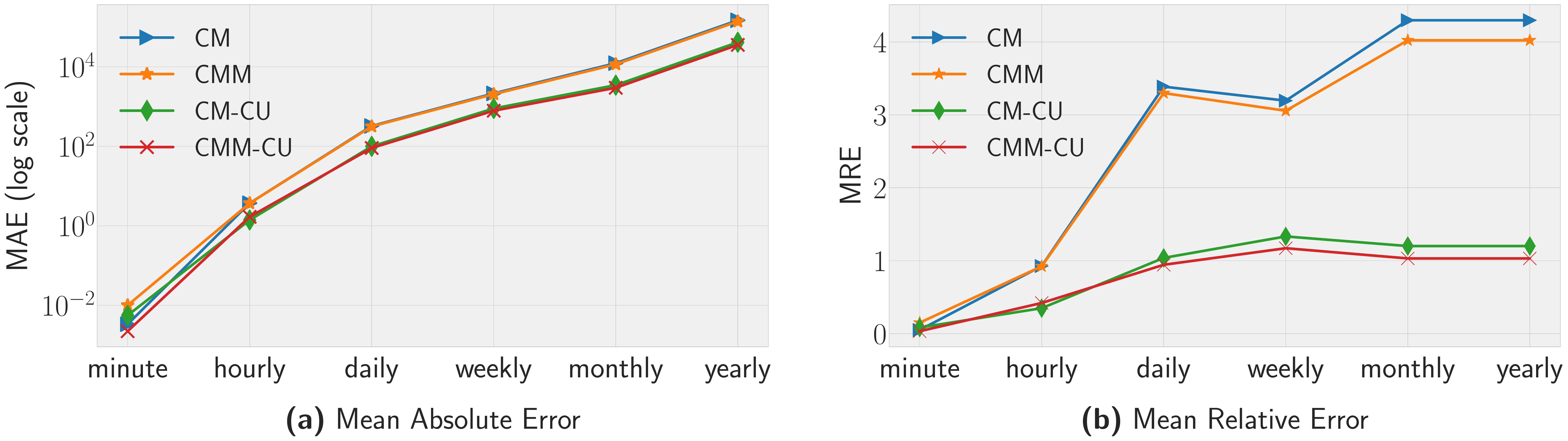}
  \caption{Occurrence tracking error using count-min sketch.}
  \Description{Occurrence tracking error using count-min sketch.}
  \label{fig:smartplug:pattern_tracking_error}
\end{figure}

Fig.~\ref{fig:smartplug:pattern_tracking} depicts the occurrence tracking results. In particular, Figs.~\ref{fig:smartplug:pattern_tracking}a-\ref{fig:smartplug:pattern_tracking}c highlight the tracking performance for pattern counts over short intervals (i.e., minute, hour, day), whereas Figs.~\ref{fig:smartplug:pattern_tracking}d-\ref{fig:smartplug:pattern_tracking}f depict the tracking results for pattern counts over longer intervals (i.e., week, month, year). Each sub-figure provides a comparison between the actual count and estimated counts obtained by each count-min sketch as indicated by the X-tick labels. For instance, the bars in Fig.~\ref{fig:smartplug:pattern_tracking}b compare the hourly occurrences (averaged over all patterns) between different sketches and the actual count. We obtain several observations:
\begin{enumerate}

  \item Approximate tracking of patterns using count-mean-min with conservative update ({\tt CMM-CU}) results in the least overestimation for all intervals from minute-by-minute to yearly tracking (Fig.~\ref{fig:smartplug:pattern_tracking}a-\ref{fig:smartplug:pattern_tracking}f). This suggests that conservative updating of the sketch and removing the estimated noise during retrieval operations significantly can improve the accuracy.

  \item Fig.~\ref{fig:smartplug:pattern_tracking_error} compares the change in {\tt MAE} and {\tt MRE} by different sketches as the interval increases from a minute to full year. As guaranteed by count-min sketch \cite{countminsketch}, the error is due to overestimation and there is no under-estimation error by any sketch. In Fig.~\ref{fig:smartplug:pattern_tracking_error}a, we observe that {\tt MAE} has a linear growth rate on a logarithmic scale, which means it actually grows exponentially with the interval length. The exponential growth is exhibited by all the sketches in Fig.~\ref{fig:smartplug:pattern_tracking_error}a. On the other hand, {\tt MRE} provides a different picture from {\tt MAE} as shown in Fig.~\ref{fig:smartplug:pattern_tracking_error}b. It shows that {\tt CM-CU} and {\tt CMM-CU} achieve superior accuracy compared to the remaining two sketches, with {\tt CMM-CU} as the best among all sketches which can be verified from Fig.~\ref{fig:smartplug:pattern_tracking}.

  \item We observe that {\tt MAE} mounts as the interval length increases. This is because when we aggregate data from one sketch to another, the error is also aggregated. For instance, the hourly estimates returned by the sketch for a pattern already contain overestimation errors. When we add them together to get the pattern's daily estimate, we implicitly add these individual hourly overestimation errors. Similarly, the weekly estimate of a pattern contains errors from both the hourly and daily estimates, and so on.

  \item Intuitively, increasing $M$ and $K$ reduces the overestimation error as already shown in \cite{countmin_improvement1} and \cite{chau_distrack}. In our cases, however, the values of $M$ and $K$ are dictated by the memory and processing capabilities of the smart plug prototype.

\end{enumerate}
From these observations, the setting {\tt CMM-CU} is shown to be desirable for tracking in the smart plug.

\subsection{Computational and Energy Overheads} \label{subsec:eval:overhead}

A key performance metric of the smart plug prototype is the computational overheads. Table~\ref{tab:smartplug:processing_overhead} lists the computational overhead for various tasks performed by the smart plug prototype. For each task, the table provides the running time required by the smart plug prototype to execute the task in increasing order of complexity. In the case of classification of known classes of patterns, for instance, the running time of the smart plug grows sub-linearly with the data length. For the remaining two tasks (i.e., autocorrelation and approximate entropy), the growth is slightly more than linear. Notably, the last column represents the maximum complexity of the given task. Overall, the smart plug prototype still performs most tasks efficiently in a timely manner. We note that the first three tasks listed in the table are related to pattern detection and classification. We have also included the computational overhead calculations for occurrence tracking. We note that each update operation requires computing only a small number of hash functions and basic arithmetic, which takes very small processing time.

\begin{table}[htbp!]
  \centering
  \scalebox{0.95}{
    %  \setlength\tabcolsep{9pt} % default value: 6pt
    %\begin{tabular}{@{} c @{}|@{} c @{}|@{} c @{}|@{} c @{}|@{} c @{}|@{} c @{}} \hline \hline
    \begin{tabular}{l|c|c|c|c|c} \hline \hline
      \multirow{2}{*}{Autocorrelation}     & Lag ($k$)                      & 50  & 100 & 200  & 400  \\ \cline{2-6}
                                           & Running Time (milliseconds)    & 24  & 88  & 344  & 1351 \\ \hline
      \multirow{2}{*}{Classification}      & Data Length ($N$)              & 200 & 400 & 800  & 1600 \\ \cline{2-6}
                                           & Running Time (milliseconds)    & 529 & 711 & 1085 & 1564 \\ \hline
      \multirow{2}{*}{Approximate Entropy} & Sliding Window Length ($\phi$) & 60  & 120 & 240  & 480  \\ \cline{2-6}
                                           & Running Time (milliseconds)    & 60  & 214 & 722  & 2310 \\ \hline \hline
      \multirow{2}{*}{ Count-min Sketch Updates} & {Number of counters in sketch ($M$)} & {100}  & {200} & {400}  & {600}  \\ \cline{2-6}
                                           & {Running Time (milliseconds) }   & {7}  & {9} & {10}  & {13} \\ \hline \hline
    \end{tabular}}
  \caption{Computational overheads of processing in the smart plug prototype.}
  \Description{Computational overheads of processing in the smart plug prototype.}
  \label{tab:smartplug:processing_overhead}
\end{table}

{
Another key performance metric of the smart plug prototype is the memory space requirements. Table~\ref{tab:smartplug:momory_overhead} details the overall memory space requirements as well as the portions of the memory space dedicated to pattern classification and occurrence tracking. In the table, the program (flash) memory space refers to the code segment (i.e., all executable instructions) and data (RAM) contains global and static variables (both initialized and uninitialized). Please note that the memory consumption by occurrence tracking is the combined memory space by all sketches (from minute-sketch to yearly-sketch), where the size of each sketch is set to $M=500$ as already mentioned in Section~\ref{subsec:eval_occurrence_tracking}.

\begin{table}[htbp!]
  \centering
  \scalebox{0.95}{
    \begin{tabular}{>{}l|>{}c|>{}c} \hline \hline
      Category                                    & Program (Flash)          & Data (RAM)              \\ \hline \hline
      Total available memory space                     & 1044 kB (1 MB)           & 82 kB                   \\ \hline
      Overall memory space by smart plug    & 611 kB (58.9\% of total) & 43 kB (52.5\% of total) \\ \hline
      Memory space by pattern detection and &                          &                         \\ classification & 338 kB (32.4\% of total) & 33 kB (40.3\% of total) \\ \hline
      Memory space by occurrence tracking   &                          &                         \\ ($M = 500$)      & 273 kB (26.5\% of total) & 10 kB (12.2\% of total) \\ \hline \hline
    \end{tabular}
  }
  \caption{ Memory space requirements of the smart plug prototype.}
  \Description{Memory space requirements of the smart plug prototype.}
  \label{tab:smartplug:momory_overhead}
\end{table}
}

\begin{table}[htbp!]
  \centering
  \scalebox{0.95}{
    \begin{tabular}{l | c } \hline \hline
      Mode                                           & Power Consumption (@3.3V) \\ \hline \hline
      Power Off                                      & 0.5 $\mu$A / 1.6 $\mu$W   \\ \hline
      Deep Sleep                                     & 10 $\mu$A / 33 $\mu$W     \\ \hline
      Algorithms                               & 15 mA / 49.5 mW           \\ \hline
      Algorithms + Current Sensor              & 55 mA / 181.5 mW          \\ \hline
      Algorithms + Web Server + Current Sensor & 185 mA / 610.5 mW         \\ \hline
    \end{tabular}}
  \caption{Energy overhead measurement of smart plug prototype.}
  \Description{Energy overhead measurement of smart plug prototype.}
  \label{tab:testbed:overhead}
\end{table}

The last key performance metric of the smart plug prototype is the energy consumption in hardware prototype. Note that the Arduino platform consumes very little energy, as compared to other platforms (e.g., Raspberry Pi). Table~\ref{tab:testbed:overhead} lists the different operating modes and their respective energy consumption measurement. The smart plug prototype only consumes larger power, when the web server is invoked (which is responsible for data visualization and user configurations). Overall, the smart plug prototype consumes far less power than the connected appliances.

%% file: related.tex
\section{Related Work} \label{sec:smartplug:related}

This work is related to the existing literature of classification of appliance types and tracking of appliance behavior.  There are two traditional approaches for the classification tasks of appliances: non-intrusive methods \cite{nilm_survey} and intrusive methods. Non-intrusive methods rely on passive measurements of power consumption of the entire apartment and disaggregating the data to identify individual appliances. On the other hand, intrusive methods involve active measurements of power consumption of individual appliances using special power meters or smart plugs. For non-intrusive methods, an experimental method for monitoring the power consumption data of a large number of appliances and extracting a small number of common characteristics is provided in \cite{appliance_modeling}. The common characteristics are then used to derive a small number of model types that describe the power usage patterns of common appliances. Another related study applies statistical and machine learning methods for automatic modeling of appliances from power consumption data  \cite{nilm}. Their method is to automatically model the appliance types and usage patterns, and then to fit a function or a distribution for the power consumption data that best describes the appliance behavior. For appliances with linear power consumption (i.e., whose current draw follows a sinusoidal curve), they use non-linear least squares method to a fit a function onto the data. For non-linear appliances, they fit a probability distribution using \emph{Maximum Likelihood Estimation} method. 
Some general surveys of processing data from smart meters and energy sensors are provided in \cite{8322199} and \cite{7548112}, respectively. However, both surveys do not cover timely processing on a smart plug.

Most of the prior studies of appliance classification are based on prior training using pre-recorded data in offline mode and do not consider resource-constrained platforms.
For example, the authors in \cite{doi:10.1177/0142331212460883} propose a technique for the classification of energy consumption patterns in industrial manufacturing systems. Their paper, however, does not apply to timely classification on end-node devices. 
A system that collects the information about energy price and makes use of a series of intelligent smart plugs, connected to each other through WiFi network, in order to optimize home energy consumption according to the user preferences is proposed in \cite{blanco2017electricity}. However, their work requires the smart plugs to communicate with each other and does not address the classification and tracking problems by resource-constraint IoT devices in a stand-alone manner, which has been addressed explicitly in our paper.

A recent study that has implemented in embedded systems is presented in \cite{auto_plug}, called AutoPlug, which can perform timely appliance classification and identification of new appliances using time series analysis and machine-learning. 
%AutoPlug is based on the Raspberry Pi 2 platform, which is more powerful but also susceptible to power disruptions than Arduino platform. 
To detect if an appliance is new or moved from one smart plug to another, AutoPlug matches the current active period with the previous one by comparing the Dynamic Time Warping distance between them or by fitting some common appliance models in each period and then comparing the fitted models to determine if device switch has occurred. AutoPlug gave around 90\% accuracy for identification of appliances and over 90\% accuracy in detecting new appliance presence and appliance changing from one smart plug to another. However, there are several key differences between AutoPlug and our work:

\begin{enumerate}
    \item AutoPlug is based on Raspberry Pi 2 that has more powerful processor and larger memory size, while our smart plug is based on Arduino micro-controller platform with much less processing power and smaller memory space. Also, Raspberry Pi runs Linux, which is not suitable for disruptible systems like smart plugs that can be unplugged from time to time. While we have to overcome the limited functionality of operating systems and compact memory space in Arduino, our smart plug is more resilient in dynamic environments. Moreover, Raspberry Pi supports pre-existing powerful scientific computing libraries for data processing and visualization, as well as modules for performing curve fitting. In our smart plug, on the other hand, we had to implement the required modules/libraries from the scratch in an efficient way to satisfy the memory and space constraints.

    \item AutoPlug is not a smart plug, but a service deployed in a wireless gateway that communicates with existing smart plugs in a house. AutoPlug assumes that the smart plugs are capable of recording and wirelessly transmitting their power consumption in real time to the gateway where AutoPlug is deployed. Our smart plug fully functions in a stand-alone manner without relying on an external gateway. 

    \item AutoPlug requires initial training from existing device energy usage traces for classification. In their evaluation, they trained the classifier on 13 different device types. Our smart plug system, on the other hand, does not require prior training.

    \item For change detection, AutoPlug employs an overly simplistic strategy which solely decides change detection based on whether or not the appliance is consuming energy. Our system, on the other hand, uses a sophisticated technique based on approximate entropy which is a proven method for detecting irregularities (i.e., changes) in data.

    \item For appliance identification, AutoPlug uses Dynamic Time Warping and curve fitting. In curve fitting, AutoPlug always fits the logarithmic growth model to the active power segment. Our system, on the other hand, uses curve fitting and online clustering for appliance identification. In curve fitting, it fits six different models (see Section~\ref{subsec:smartplug:classif_known_patterns}) and chooses the model with the best fit among the six models, thus enabling our system to detect a wider range of appliances. 

\end{enumerate}

In addition to the application of smart plugs, this paper studies generic systems for continuous signal monitoring and tracking. We draw upon the existing literature of streaming data algorithms. The basic idea of streaming data algorithms is to make use of randomized data structures that are able to amortize the worst-case inputs with high probability. An example of streaming data algorithms is clustering, which is used for classification of signals. A number of clustering algorithms have been proposed for streaming data, which can be divided into two classes: (i) algorithms that summarize the streaming data in an online manner, and generate clusters from the summary in offline mode, (ii) algorithms that are able to cluster the streaming data entirely in an online manner without offline processing. Our smart plug focuses on the later class as  the ability to provide online clustering, without offline processing, should be the most critical requirement of IoT. An extensive survey of both types of algorithms can be found in \cite{streaming_review2}.

We briefly survey the related clustering algorithms in the literature. For example, \cite{clustering_paper1} present a streaming algorithm that clusters large streaming data using a small amount of memory, by processing the data in a small number of passes. A clustering algorithm for parallel streams of continuously evolving time series data is proposed in \cite{clustering_paper2}. The streaming data is grouped based on the similarity computed using an online variant of the k-Means clustering algorithm. Another study \cite{clustering_paper3} presents a technique for clustering of time series data based on structural features instead of distance metrics. The purpose of using structural features (e.g., correlation, skewness, seasonality, etc.) is for dimensionality reduction and reduction of sensitivity due to missing data. The best set of features is determined and used as input to various algorithms like neural networks, self-organizing map, or hierarchical clustering to can potentially provide meaningful clusters. Comprehensive reviews of various experimental methods for clustering streaming data and their applications are provided in \cite{streaming_review1,clustering_review}.

In this work, we adopt streaming data algorithms for tracking the occurrences of signal patterns in embedded systems. The existing studies of streaming data algorithms focus on a number of counting  problems, such as heavy hitters \cite{Cormode_heavyhitter}, frequency moments \cite{superspreader,flowentropy,bitmap}, which have been applied to the applications in database systems and network traffic measurement.  However, the specific application to cyber-physical systems for tracking continuous signals in this paper is novel.
Beyond tracking occurrences, tracking temporal correlations using streaming data algorithms was studied in \cite {chau_distrack}, which presents online traffic tracking algorithms for network management considering both space and speed constraints. It addresses the challenge of balancing the trade-off between space and accuracy and caters to distributed deployment in large networks with heterogeneous local storage space constraints.

Finally, a preliminary version of this work appeared in \cite{aftabAPsys}, which is now extended to include a more detailed experimental evaluation study and the function of occurrence tracking. See Appendix for detailed descriptions.

%% file: concl.tex
\section{Conclusion and Future Work} \label{sec:smartplug:concl}

This work presented a timely processing smart plug system design and an implemented prototype that can perform timely monitoring and tracking of the appliance power consumption behavior such as timely detection, classification, and clustering of appliance behavior and patterns. These monitoring features offer many benefits to researchers, such as automated demand response, appliance localization, etc. The smart plug prototype is developed using low-cost Arduino open hardware platform. In general, our timely classification and tracking systems can be applied to a variety of near real-time processing smart sensors for wearable, biomedical, and environmental monitoring applications. In the following, we summarize some  important lessons learned in our study of enabling smart plugs for timely classification of appliance types and tracking of appliance: 

\begin{itemize}
	
\item	Offline processing can provide accurate results at the expense of latency, which creates a problem for timely applications of IoT. Online processing, on the other hand, can provide timely processing with an impact on accuracy. However, our study shows that effective design and implementation of online processing algorithms, such as learning and classifying patterns of continuous signals, can produce results close to offline processing. 
	
 \item 	Memory space constraint is another issue of real-time IoT devices. This study demonstrated the use of memory-efficient tracking can effectively accommodate small local memory space while providing satisfactory performance.

\end{itemize}

In the future, we plan to use the smart plug system for implementing the functionality to identify the context of events by finding correlations among the events of multiple smart plugs. Our system can also be extended to support additional features, such as real-time detection of anomalous appliance behavior and effective power allocation. 

%% file: append.tex
\section*{Table of Key Notations}

\begin{table}[htbp!]
  \centering
  \scalebox{0.85}{
  \begin{tabular}{  >{\color{black}}l | >{\color{black}} l } \hline \hline
    %\begin{tabular}{l | p{11cm} } \hline \hline
    Parameter                   & Definition                                                                                                                                          \\ \hline \hline
    $x[t_1:t_2]$                & A segment of signal (i.e., time series) from time $t_1$ to $t_2$                                                                                      \\ \hline
    ${\cal S}$                  & A non-overlapping segmentation of time                                                                                                              \\ \hline
    ${\cal C}$                  & The set of observed patterns (i.e., cluster of patterns)                                                                                                \\ \hline
    $c_i$                       & A canonical signal (typically cluster center) to represent a cluster of observed patterns                                                             \\ \hline
    %${\sf d}\big(x[t_1: t_2], c_i\big)$ & Distance metric between $x[t_1: t_2]$ and $c_i$                                                        \\ \hline
    ${\sf d}\big(x_1, x_2\big)$ & Distance metric between segments $x_1$ and $x_2$                                                                                                    \\ \hline
    $k$                         & The number of clusters                                                                                                                                  \\ \hline
    $N_i^{\cal T}$              & The true total number of occurrences of pattern $i \in {\cal C}$ in a certain epoch of time ${\cal T}$                                              \\ \hline
    $\widehat{N}_i^{\cal T}$    & The estimated total number of occurrences at the end of ${\cal T}$                                                                                  \\ \hline
    $\epsilon$ and $\delta$     & Parameters for controlling the trade-off between accuracy and memory size                                                                           \\ \hline
    %    $c$                                 & Memory size                                                                                            \\ \hline
    $N$                         & Total length of the time series $x$                                                                                                                 \\ \hline
    $m$                         & Positive integer representing the length of a sub-sequence                                                                                          \\ \hline
    $\theta$                    & Positive integer representing the similarity threshold between a pair of sub-sequences                                                              \\ \hline
    $s^m$                       & The set of sub-sequences of length $m$ in the time series. $m$ is used internally by the \\ & approximate entropy ({\tt ApEn}) algorithm for comparison.
    \\ \hline
    $\phi$                      & Length of the sliding window over which {\tt ApEn} is calculated                                                                                    \\ \hline
    %$C_i^m(\phi)$               & The fraction of sub-sequences in $s^m$ that are similar to $x[i:i+m-1]$ for a sliding window of length $\phi$ \\ \hline
    $C_i^m$                     & The fraction of sub-sequences in $s^m$ that are similar to $x[i:i+m-1]$                                                                             \\ \hline
    $H$                         & Computed {\tt ApEn} values                                                                                                                          \\ \hline
    $E$                         & The number of edges detected by {\tt ApEn}                                                                                                              \\ \hline
    $\eta$                      & The number of most recent {\tt ApEn} values over which the online window operates                                                                   \\ \hline
    $\omega$                    & Edge separation threshold                                                                                                                           \\ \hline
    $z_t$                       & Newly received segment of the signal                                                                                                                \\ \hline
    ${\cal Z}$                  & The set of cluster centers corresponding to the set of clusters ${\cal C}$                                                                          \\ \hline
    $d^*$                       & The minimum inter cluster distance in ${\cal C}$                                                                                                    \\ \hline
    ${|{\cal C}_i|}$            & The total number of elements in the $i$-th cluster                                                                                                  \\ \hline
    $\alpha$                    & The relative weight of the new segment $z_t$, which provides an effect of exponential smoothing                                                     \\ \hline
    $h(.)$                      & Hash function                                                                                                                                       \\ \hline
    $K$                         & The number of hash function in count-min sketch                                                                                                         \\ \hline
    $M$                         & The number of counters in count-min sketch                                                                                                              \\ \hline
    $C_{i, h_t^i}$              & The $h_t^i$-th counter in the $i$-th row of the sketch                                                                                              \\ \hline
    \hline
  \end{tabular}
  }
  \caption{List of key notations and their definitions.}
  \Description{List of key notations and their definitions.}
  \label{tab:append:params}
\end{table}